\documentclass[aps,prb,showpacs,twocolumn,superscriptaddress]{revtex4-1}
\usepackage{graphicx,amsmath,amssymb}
\usepackage[usenames]{color}
\usepackage{indentfirst}
\usepackage{float}
\usepackage[T1]{fontenc}
\usepackage[colorlinks=true, citecolor=blue, urlcolor=blue, linkcolor=blue ]{hyperref}
\hypersetup{breaklinks=true}
\begin{document}

\bibliographystyle{apsrev4-1}

\title{Altermagnetism in exactly solvable model: the Ising-Kondo lattice model}
\author{Miaomiao Zhao}
\affiliation{Key Laboratory of Quantum Theory and Applications of MoE $\&$ School of Physical Science and Technology, Lanzhou University, Lanzhou 730000, People Republic of China}
\affiliation{Lanzhou Center for Theoretical Physics, Key Laboratory of Theoretical Physics of Gansu Province, Lanzhou University, Lanzhou 730000, People Republic of China}
\author{Wei-Wei Yang}
\affiliation{Beijing National Laboratory for Condensed Matter Physics and Institute of Physics, Chinese Academy of Sciences, Beijing 100190, China}
\author{Yin Zhong}
\email{zhongy@lzu.edu.cn}
\affiliation{Key Laboratory of Quantum Theory and Applications of MoE $\&$ School of Physical Science and Technology, Lanzhou University, Lanzhou 730000, People Republic of China}
    \affiliation{Lanzhou Center for Theoretical Physics, Key Laboratory of Theoretical Physics of Gansu Province, Lanzhou University, Lanzhou 730000, People Republic of China}
\begin{abstract}
 Altermagnet (AM), a recently identified class of collinear magnet, has garnered significant attention due to its unique combination of zero net magnetization and spin-split energy bands, leading to a variety of novel physical phenomena. Using numerically exact lattice Monte Carlo simulations, we investigate AM-like phases within the Ising-Kondo lattice model which is commonly employed to describe  heavy-fermion materials. By incorporating an alternating next-nearest-neighbor hopping (NNNH) term, which arises from the influence of non-magnetic atoms in altermagnetic candidate materials, our results reveal key signatures of AM-like states, including spin-splitting quasiparticle bands and spectral functions, and demonstrate that $d$-wave AM remains stable across a broad range of interaction strengths, doping levels, NNNH amplitudes and temperatures, highlighting its robustness. Furthermore, through an analysis of non-magnetic impurity effects, we further confirm the $d$-wave symmetry of the AM phase. These findings establish a solid theoretical foundation for exploring AM-like phases in $f$-electron compounds, paving the way for future investigations into their exotic magnetic and electronic properties.
\end{abstract}

\maketitle
\section{Introduction}

An emerging class of collinear magnet, recently named after altermagnet (AM), \cite{Smejkal2022,PRXSmejkal2022,Mazin2022} has been discovered. \cite{Ahn2019,Phys.Soc.JpnHayami2019,Smejkal2020,PRBHayami2020,Yuan2020,PRLGonzalez-Hernandez2021,Mazin2021,Bai2022} Like traditional antiferromagnet, AM exhibits zero net magnetization and consists of at least two spin sublattices. However, they are distinguished by their energy bands, which display spin splitting driven by the breaking of time-reversal or rotation symmetry. This behavior aligns more closely with the ferromagnet than with the antiferromagnet, setting AM apart as a unique magnetic phase. The distinctive properties of AM give rise to a variety of novel physical phenomena, including spin currents,\cite{Phys.Soc.JpnHayami2019, Bai2023, PRBCui2023,AdvSciGuo2024,Sicheler2025} spin-splitting torque effects,\cite{Bai2022,PRLKarube2022} and anomalous Hall effect, \cite{Smejkal2020,Gonzalez2023,PRLSato2024,PRBAttias2024,NatCommunReichlova2024} making them a focal point of intense research interest. Several candidate materials for AM have been extensively investigated, including RuO$_{2}$, MnTe, MnF$_{2}$, CrSb and even La$_{2}$CuO$_{4}$, the parent compound of the high-$T_{c}$ cuprate superconductor.\cite{Krempasky2024,Fedchenko2024,Bai2022,Feng2022,Gonzalez2023,Bai2023,Hariki2024,Lee2024,Bhowal2024PRX,Li2024PRB,Osumi2024,Yang2024} Notably, spin-dependent band splitting with $d$- or $g$-wave symmetry has been experimentally verified in RuO$_{2}$, MnTe and CrSb.\cite{Krempasky2024,Fedchenko2024,Lee2024,Osumi2024,Yang2024}

 AM, characterized by spin-polarized band structures with broken time-reversal symmetry, has emerged as a versatile platform for exploring a wide range of phenomena in condensed matter physics,\cite{Brekke2023,Mland2024,Bose2024,Maier2023,Das2024,Leeb2024} particularly in many-body systems involving topology, superconductivity and Kondo effect. \cite{Chakraborty2023,Antonenko2025,DelRe2025,Zhu2023,PRBDiniz2024,ArxivLee2023,PRLOuassou2023,Brekke2023,PRBFernandes2024,PRBCheng2024} While topological properties and superconductivity in AM have been extensively studied, the Kondo effect, which is closely associated with the heavy fermion materials, has received comparatively less attention. It is well known that heavy fermion materials are strongly correlated electron systems, in which localized magnetic moments, arising from partially filled $f$-electron orbitals, interact with conduction electrons.\cite{Hewson1993} Two dominant mechanisms govern their properties: the Kondo effect, which screens local spins through interactions with conduction electrons, and the Ruderman-Kittel-Kasuya-Yosida (RKKY) interaction, which mediates indirect exchange between localized moments. In the context of AM, the heavy fermion material Ce$_4$Sb$_3$ has been explored using first-principles calculations and analysis of tight-binding model. Those studies reveal that the AM state in Ce$_4$Sb$_3$ hosts topological phases exhibiting exotic phenomena, including spin-splitter torque and pronounced nonlinear transport effects.\cite{arXivHe2024} Additionally, strong parity breaking and anisotropic symmetry lowering in spin-polarized, time-reversal symmetry-broken Fermi surfaces have been observed in another heavy fermion material, CeNiAsO. \cite{arXivHellenes2023}  Unlike the conventional even-parity AM with $d$- or $g$-wave symmetry, CeNiAsO displays $p$-wave symmetry, earning it the designation of a $p$-wave AM.
 Furthermore, the Kondo effect and RKKY interaction arising from single or two impurities have been studied in altermagnetic systems.\cite{PRBDiniz2024, ArxivLee2023,PRBAmundsen2024} These investigations demonstrate that AM will significantly influence the Kondo temperature and the RKKY interaction exhibits an anisotropic oscillatory pattern, reflecting the $C_{4z}$ symmetry inherent to AM. 
 
The Kondo lattice model, which incorporates exchange interactions between itinerant and localized electrons, serves as a powerful framework for describing heavy fermion materials.\cite{Tsunetsugu,Coleman2015} By combining the antiferromagnet states in the Kondo lattice model\cite{Lacroix1979,Fazekas1991,Zhang2000,Li2015,Watanabe2007,Asadzadeh2013,Martin2008,Lenz2008} with an alternating next-nearest-neighbor-hopping (NNNH) derived from non-magnetic atoms, \cite{Das2024} the AM in heavy fermion compounds have been discussed using mean-field approach.\cite{Zhong2024} This study has revealed a remarkably rich phase diagram, encompassing distinct phases: the Kondo screening state, the AM state and a coexistent state where AM and Kondo screening coexist. However, when the NNNH term is non-zero, the Kondo lattice model becomes challenging to solve exactly, even at half-filling, using unbiased quantum Monte Carlo simulations due to the fermion minus-sign problem.\cite{Assaad1999} In addition to the general Kondo lattice model, certain heavy fermion compounds exhibit easy-axis magnetic order, hence allowing the transverse Kondo coupling to be neglected.\cite{Lohneysen2007, Si2013, Coleman2010} In such cases, the Kondo lattice model is reduced to the Ising-Kondo lattice model which is first proposed to account for the anomalously small staggered magnetization and large specific heat jump observed at the hidden order transition in URu$_2$Si$_2$.\cite{PRBSikkema1996,RevModPhysMydosh2011} In contrast to the isotropic Kondo lattice model, the Ising-Kondo lattice model can be exactly solvable by using Monte Carlo simulation for arbitrary electron fillings and lattice geometries, and the antiferromagnet states have been shown to remain stable across a broad range of parameters in the phase diagram.\cite{PRBYang2019, PRBYang2020, PRBYang2021}

Inspired by the aforementioned important progress, in this work, we introduce an exactly solved model to attack AM. It is an Ising-Kondo lattice model with an alternating NNNH on a square lattice, as illustrated in Fig.~\ref{fig:model}. Through a combination of analytical arguments and numerically exact lattice Monte Carlo (LMC) simulations,\cite{Maska2006} we uncover $d$-wave AM with spin-splitting quasiparticle energy bands in the Ising-Kondo lattice model.

 The remainder of this article is organized as follows. In Sect.~\ref{sec1}, we introduce the Ising-Kondo lattice model on a square lattice with an alternating NNNH term and elucidate why this model is exactly solvable. Sect.~\ref{sec2} presents the ground state phase diagram using Monte Carlo simulations and some analytical arguments, identifying the $d$-wave AM phase in the Ising-Kondo lattice model through its characteristic spin-splitting band structure and spin-resolved spectral function. In Sect.~\ref{sec3}, we analyze the density of states and conductivity in the AM phase, and also explore the effects of finite temperature and a non-magnetic impurity on the Ising-Kondo lattice model. Finally, a brief summary is given in Sect.~\ref{sec4}.
  

\section{Model and Exact solvability}\label{sec1}

\begin{figure}[H]
	\begin{centering}
		\includegraphics[width=0.41\textwidth]{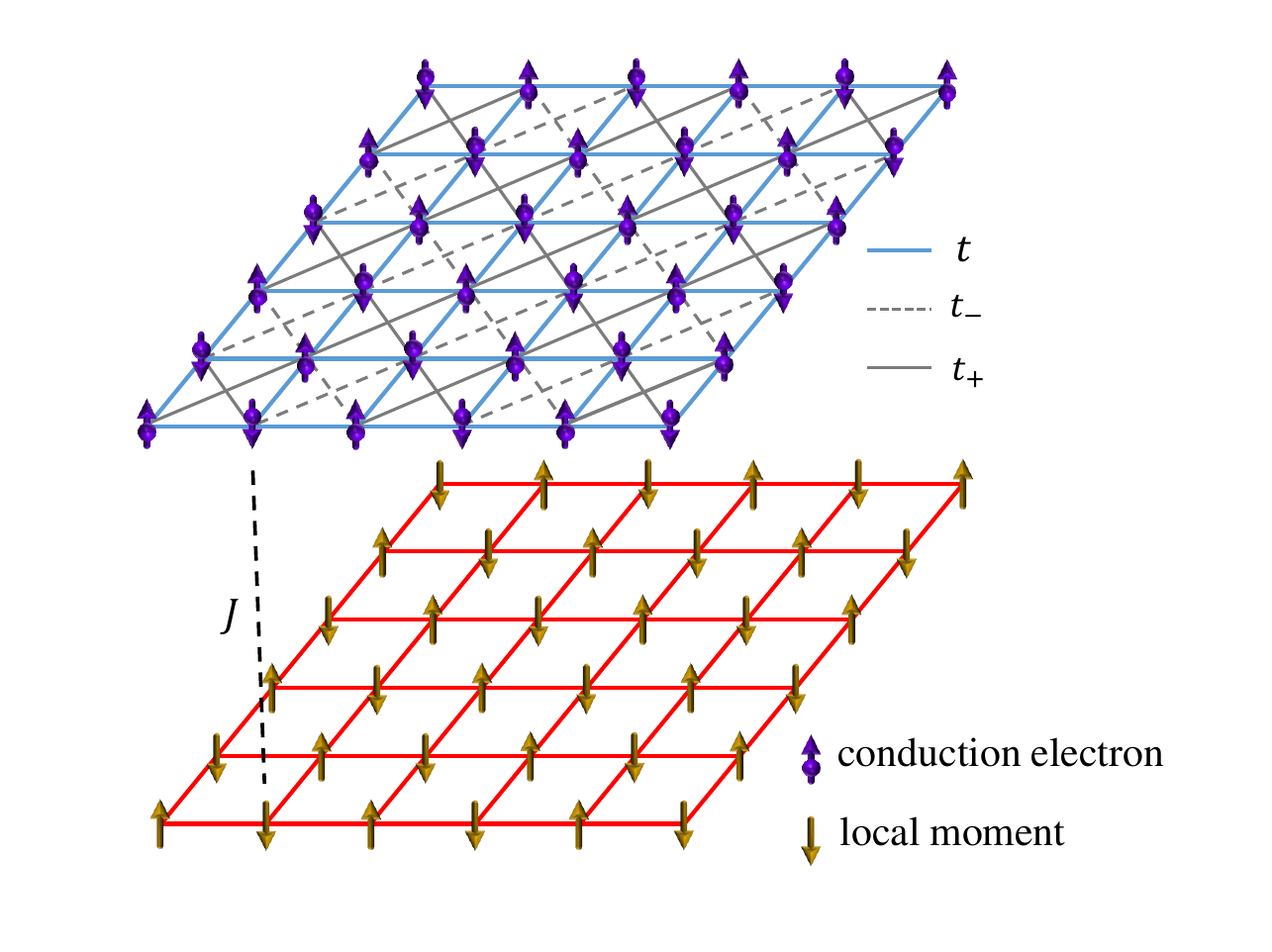}
		\par\end{centering}
	\protect\caption{
		\label{fig:model} The Ising-Kondo lattice model on a square lattice with alternating next-nearest-neighbor-hopping (NNNH), $t_{+}$ and $t_{-}$. The lower layer denotes local spin moments of $f$-electron, which interact with the conduction electron in the upper layer via only longitudinal Kondo exchange.}
\end{figure}

We consider an anisotropic limit of the Kondo lattice model, namely the Ising-Kondo lattice model on square lattice (Fig.~\ref{fig:model})
\begin{eqnarray}
	\hat{H}=-\sum_{i,j,\sigma} t_{ij}\hat{c}^{\dag}_{i \sigma}\hat{c}_{j\sigma} + \frac{J}{2}\sum_{j, \sigma}\hat{S}^{z}_{j}\sigma\hat{c}^{\dag}_{j \sigma}\hat{c}_{j \sigma } \label{HIsing}
\end{eqnarray}
where $\hat{c}_{j\sigma}^{\dagger}$ is the creation operator of conduction electron ($c$-electron) with spin flavor $\sigma =\uparrow$, $\downarrow$. $\hat{S}_{j}^{z}$ denotes the local spin moment of $f$-electron at the site $j$. $J$ is the longitudinal antiferromagnetic ($J>0$) Kondo coupling (also referred to as the Ising-Kondo coupling/interaction) between conduction electrons and localized spin moments of $f$-electrons. 
$t_{ij}$ is the hopping integral between $i$, $j$ sites for the $c$-electron. Following the approaches in Refs.~\onlinecite{Das2024} and \onlinecite{Zhong2024}, in order to induce AM in this model, nearest-neighbor-hopping (NNH), denoted by $t$, and NNNH, denoted by $t_{+}$ and  $t_{-}$, have been taken into account (shown in Fig.~\ref{fig:model}). Most importantly, AM phases emerge only when the NNNH is alternating, which breaks the spin-flip and translation symmetry or the $C_{4z}$ ($\pi/2$-rotation) symmetry of the original lattice. 

When $t_{+}\neq t_{-}$, the lattice has A and B sublattice structure and the non-interacting part of Hamiltonian, denoted by $\hat{H}_{0}$, reads
\begin{eqnarray}
	\hat{H}_{0}&=&-\sum_{i,j,\sigma} t_{ij}\hat{c}^{\dag}_{i \sigma}\hat{c}_{j\sigma} \nonumber \\ 
	&=&-t\sum_{i,\delta,\sigma}(\hat{c}_{iA\sigma}^{\dag}\hat{c}_{i+\delta,B\sigma}+\hat{c}_{i+\delta,B\sigma}^{\dag}\hat{c}_{iA\sigma})\nonumber\\
	&-&\sum_{i,\delta_{1}',\sigma}(t_{-}\hat{c}_{iA\sigma}^{\dag}\hat{c}_{i+\delta_{1}',A\sigma}+t_{+}\hat{c}_{iB\sigma}^{\dag}\hat{c}_{i+\delta_{1}',B\sigma})\nonumber\\
	&-&\sum_{i,\delta_{2}',\sigma}(t_{+}\hat{c}_{iA\sigma}^{\dag}\hat{c}_{i+\delta_{2}',A\sigma}+t_{-}\hat{c}_{iB\sigma}^{\dag}\hat{c}_{i+\delta_{2}',B\sigma}).
\end{eqnarray}
The vectors of NNH are $\delta = (\pm 1, 0)$ and $(0, \pm 1)$, while the vectors of NNNH are $\delta_{1}'=(1, 1)$, $(-1, -1)$ and $\delta_{2}' = (1, -1)$, $(-1, 1)$.

The merit of Hamiltonian Eq.~\ref{HIsing} lies in its exact solvability due to $[\hat{S}_{j}^{z},\hat{H}] = 0$. By choosing the eigenstates of $\hat{S}_j^z$ as the basis, where $\hat{S}_j^z |q_j\rangle = \frac{q_j}{2} |q_j\rangle$ with $q_j = \pm 1$, 
 the model automatically reduces to an effective free fermion model,
\begin{eqnarray}
	\hat{H}(q_{j})=\hat{H}_{0} + \sum_{j, \sigma} \frac{J\sigma}{4}q_{j}\hat{c}^{\dag}_{j \sigma}\hat{c}_{j \sigma } \label{HIsingFree}
\end{eqnarray}
Now the many-body eigenstates of the original model (Eq.~\ref{HIsing}) can be constructed via the single-particle states of the effective Hamiltonian $\hat{H}(q_{j})$ under a given configuration of the effective Ising spin $\{q_{j}\}$. Consequently, Eq.~\ref{HIsing} is exactly solvable and works as an effective spinfull Falicov-Kimball model. \cite{PRLFalicov1969} Because only local conservation of the $f$-electron spin moment is essential for reducing the system to an effective free fermion formalism, our model is solvable for arbitrary lattice geometry, spatial dimension, and electron filling, in contrast to the isotropic Kondo lattice model, where the notorious fermion minus-sign problem precludes exact numerical solution. Furthermore, the inclusion of an external magnetic field along the $z$-axis or an Ising interaction term $\sum_{i,j} J_{ij} \hat{S}_i^z \hat{S}_j^z$ does not compromise the model’s solvability. Leveraging these properties, we can utilize LMC simulation \cite{Maska2006} to solve this model numerically exactly. (The details of LMC simulation can be seen in the appendix of Ref.~\onlinecite{PRBYang2019} )

\section{The ground state}\label{sec2}
\subsection{$(t_{-}-t_{+}, n_{c})$ phase diagram} \label{subA}

\begin{figure}[H]
	\begin{centering}
		\includegraphics[width=0.42\textwidth]{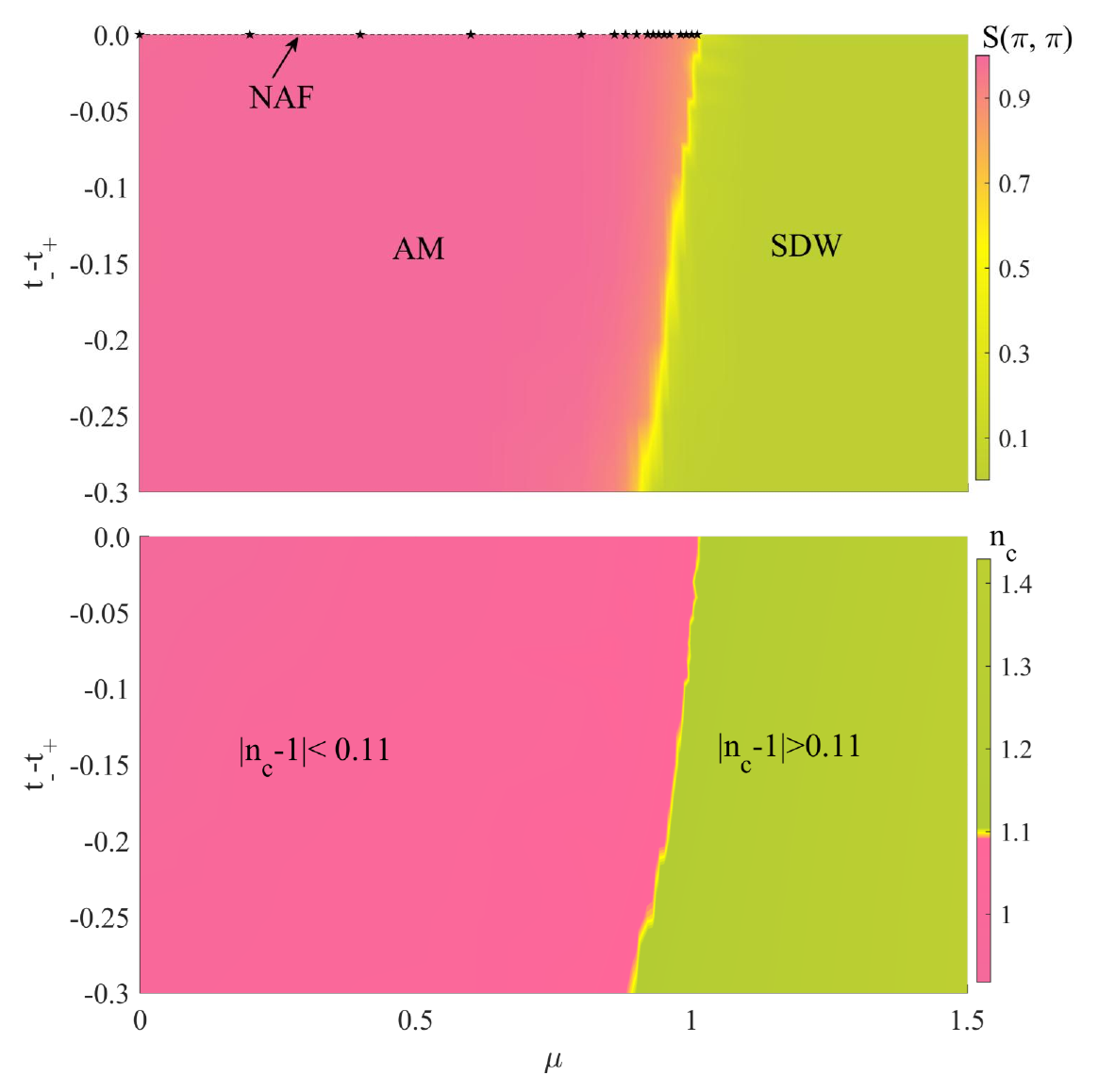}
		\par\end{centering}
	\protect\caption{
		\label{fig:phase} (a) The schematic ground state phase diagram of the Ising-Kondo lattice model as a function of NNNH $t_{-}-t_{+}$ and chemical potential $\mu$. There exist three distinct states, i.e., altermagnet (AM), normal antiferromagnet (NAF, corresponding to $t_{-}-t_{+}=0$, i.e., the dashed line with pentagram) and the spin density wave (SDW). (b) the range of electron density for part (a) with varying $t_{-}-t_{+}$ and $\mu$. Only when $n_{c}$ is in the vicinity of half filling and $t_{-}\neq t_{+}$, AM occurs. The other parameters are $J=3$ and $t_{+}$=0.3.}
\end{figure}

 The trivial Ising-Kondo lattice model, when half filled and situated on a bipartite lattice, exhibits a ground state configuration of $\{q_{j}\}$ characterized by a twofold degenerate checkerboard order $q_{j}=\pm (-1)^{j}$. This finding is consistent with the theorem proven by Kennedy and Lieb\cite{PhysicaAKennedy1986} and has also been corroborated by LMC simulation.\cite{PRBYang2019} This state, i.e., the Ising antiferromagnetic long-ranged order, is usually used to describe the localized $f$-electron spin moment. In contrast to the standard Ising-Kondo lattice model, our model incorporates an alternating NNNH term, and we also observe the checkerboard order $q_{j}=\pm (-1)^{j}$ when the electron is half-filling. (as shown in Fig.~\ref{fig:phase} (a), where the phase diagram is determined by the structure factor $S(\pi, \pi)$.) Furthermore, according to Fig.~\ref{fig:phase}, this order remains stable even with small doping. Thus, when our model is in the vicinity of half filling, the single-particle Hamiltonian of the conduction electrons in the ground state is
\begin{eqnarray}
	\hat{H}=-\sum_{i,j,\sigma} t_{ij}\hat{c}^{\dag}_{i \sigma}\hat{c}_{j\sigma} + \sum_{j, \sigma} \frac{J\sigma}{4}(-1)^{j+1}\hat{c}^{\dag}_{j \sigma}\hat{c}_{j \sigma } \label{HIsingFreeG}
\end{eqnarray}
 
In the following, we will demonstrate how this particular sublattice-dependent diagonal hopping combined with the antiferromagnetic order generates AM.

Due to the translation symmetry of the unit cell of lattice, we can separately perform a Fourier transform on the operators for the A and B sublattices in Hamiltonian (Eq.~\ref{HIsingFreeG}), i.e., $\hat{c}_{jA\sigma}=\frac{1}{\sqrt{N_s}}\sum_{k}e^{ikR_{j}}\hat{c}_{Ak\sigma}$, $\hat{c}_{jB\sigma}=\frac{1}{\sqrt{N_s}}\sum_{k}e^{ikR_{j}}\hat{c}_{Bk\sigma}$, where $N_{s}$ is the number of unit cells and the number of sites $N$ should be $2N_{s}$. Then the Hamiltonian can be expressed in momentum $k$ space:
\begin{eqnarray}\label{H_gk}
	\hat{H}(k)&=&\sum_{k,\sigma}(\varepsilon_{k}\hat{c}_{Ak\sigma}^{\dag}\hat{c}_{Bk\sigma}+h.c.+(\varepsilon_{k}^{AA}+\frac{J\sigma}{4})\hat{c}_{Ak\sigma}^{\dag}\hat{c}_{Ak\sigma} \nonumber \\
	&+&(\varepsilon_{k}^{BB}-\frac{J\sigma}{4})\hat{c}_{Bk\sigma}^{\dag}\hat{c}_{Bk\sigma})\nonumber\\ 
	&=& \sum_{k,\sigma} \left( \begin{array}{lr}\hat{c}_{Ak \sigma}^{\dagger} &\hat{c}_{Bk\sigma}^{\dagger}  \end{array}\right) \left( \begin{array}{cc} \varepsilon_{k\sigma}^{AA} &\varepsilon_{k} \\ \varepsilon_{k}&\varepsilon_{k\sigma}^{BB} \end{array}\right) \left( \begin{array}{l}\hat{c}_{Ak \sigma} \\ \hat{c}_{Bk\sigma}  \end{array}\right)\nonumber\\
\end{eqnarray} 
where $\varepsilon_{k\sigma}^{AA}=\varepsilon_{k}^{AA}+J\sigma/4$, $\varepsilon_{k\sigma}^{BB}=\varepsilon_{k}^{BB}-J\sigma/4$, $\varepsilon_{k}^{AA}=-2t_{-}\cos(k_{x}+k_{y})-2t_{+}\cos(k_{x}-k_{y})$, $\varepsilon_{k}^{BB}=-2t_{+}\cos(k_{x}+k_{y})-2t_{-}\cos(k_{x}-k_{y})$ and $\varepsilon_{k}=-2t[\cos(k_{x})+\cos(k_{y})]$. $\varepsilon_{k}^{AA}$ and $\varepsilon_{k}^{BB}$ denote intra-sublattice hopping energies, while $\varepsilon_{k}$ represents inter-sublattice hopping energy. The summation over momentum $k$ in the above equation is the magnetic Brillouin zone or reduced Brillouin zone due to the A and B sublattice structure, which is determined by the area enclosed by $k_{y} = -k_{x} \pm \pi$ and $k_{y} = k_{x} \pm \pi$. 

In order to fix the electron density at a given filling, we must take the chemical potential $\mu$ into account. Then we can solve the above Hamiltonian and obtain four spectra:
\begin{eqnarray}
	E_{k\uparrow \pm}=\frac{1}{2}\left[\varepsilon_{k}^{AA}+\varepsilon_{k}^{BB}-2\mu\pm \sqrt{4\varepsilon_{k}^{2}+(\varepsilon_{k}^{AA}-\varepsilon_{k}^{BB}+\frac{J}{2})^{2}}\right] \nonumber \\
	E_{k\downarrow \pm}=\frac{1}{2}\left[\varepsilon_{k}^{AA}+\varepsilon_{k}^{BB}-2\mu\pm \sqrt{4\varepsilon_{k}^{2}+(\varepsilon_{k}^{AA}-\varepsilon_{k}^{BB}-\frac{J}{2})^{2}}\right] \nonumber
\end{eqnarray}
When the antiferromagnetic order exists $(J\neq0)$, the spectra of electrons are spin-splitting, consisting with the requirements of AM. Additionally, as an isotropic NNNH term ( $t_{+}=t_{-}=t_{1}$) is considered, $\varepsilon_{k}^{AA}=\varepsilon_{k}^{BB}=-4t_{1}\cos(k_{x})\cos(k_{y})$ and the spin splitting in the electron spectra will vanish. Hence, alternating NNNH is also a requirement of AM in our model.
\begin{figure}[H]
	\begin{centering}
		\includegraphics[width=0.4\textwidth]{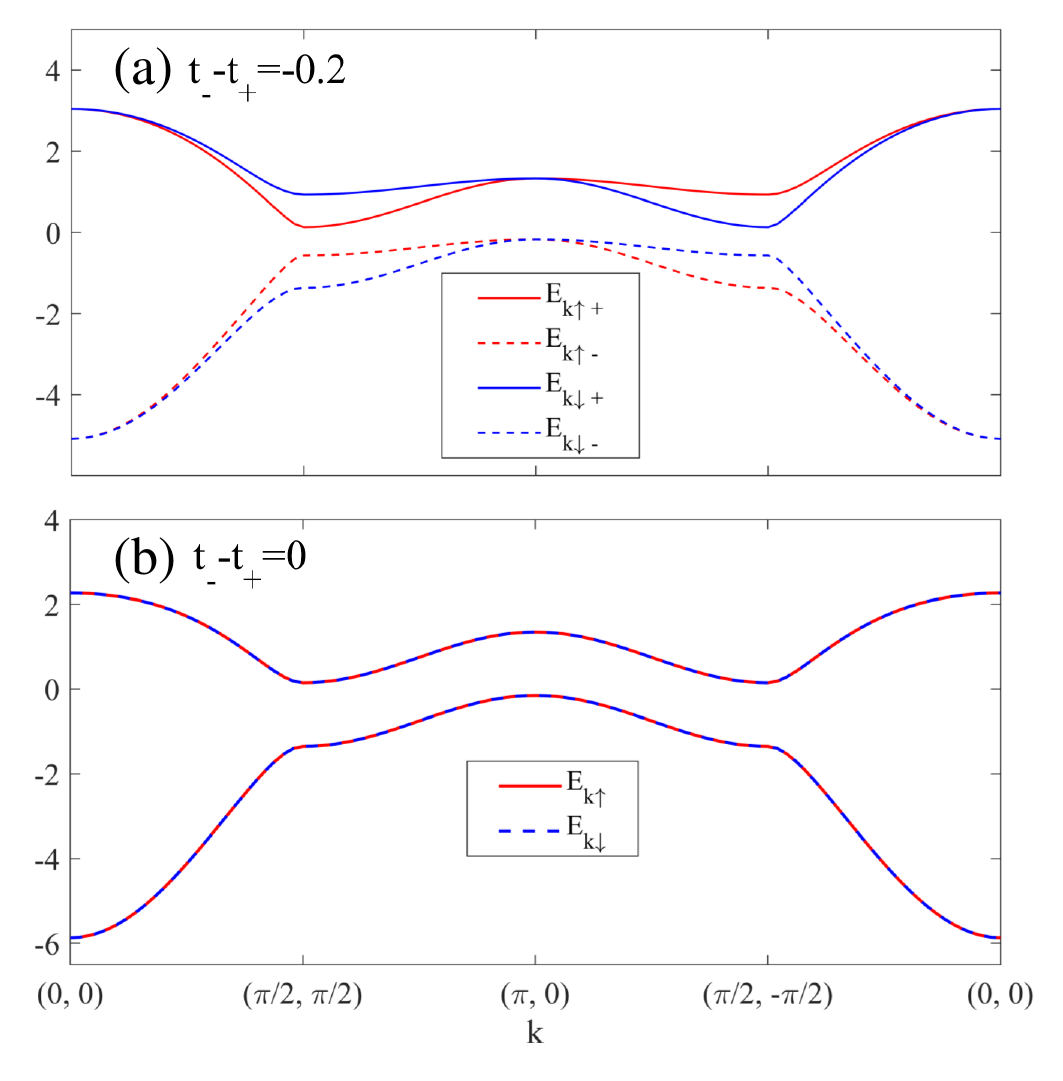}
		\par\end{centering}
	\protect\caption{ (a) Spin-splitting bands in the AM state ($t_{-}-t_{+}=-0.2$). The $C_{4z}$ symmetry of bands indicates a d-wave AM state. (b) Spin-degenerated bands in the NAM state ($t_{-}-t_{+}=0$). The other parameters are $J=3$, $t_{+}=0.3$ and $n_{c}=1$.
		\label{fig:disp} }
\end{figure}

\begin{figure}[H]
	\begin{centering}
		\includegraphics[width=0.42\textwidth]{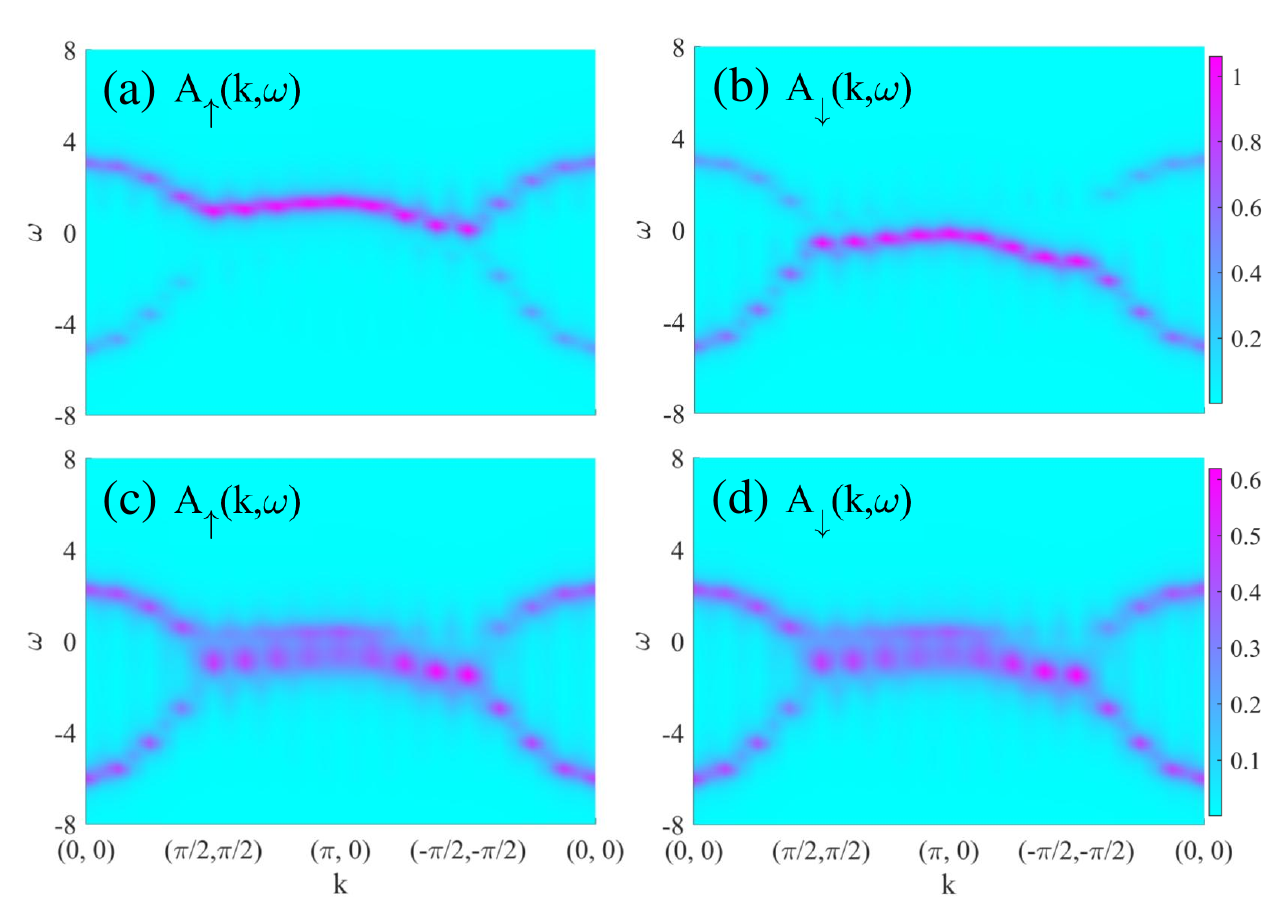}
		\par\end{centering}
	\protect\caption{
		\label{fig:Aktn0p1} The spectral function $A(k,\omega)$ for $t_{-}-t_{+}=-0.2$. (a) and (b) correspond to the cases for spin-up and spin-down, respectively, with $\mu=0.22$ (half filling) in AM; (c) and (d) correspond to the cases for $\mu=1.1$ (far away from half-filling) in SDW. The other parameters are $t_{+}=0.3$ and $J=3$. }
\end{figure}

To illustrate the band splitting caused by AM more intuitively, we present Fig.~\ref{fig:disp}(a). ($t_{-}-t_{+}=-0.2$ and the electron is half filled.) We observe spin-splitting energy bands, which is the characteristics of AM. For comparison, we also plot the dispersion curve for $t_{-}-t_{+}=0$ (shown in Fig.~\ref{fig:disp}(b)). From Fig.~\ref{fig:disp}(b), no spin splitting of bands is observed, and hence the ground state is in a normal antiferromagnetic order due to the equal NNNH strength, which agrees with the phase diagrams in Fig.~\ref{fig:phase} and emphasizes the key role of the staggered NNNH for AM. Furthermore, the bands in the AM state exhibit $C_{4z}$ symmetry (i.e., under $(k_{x},k_{y})\to (-k_y,k_x)$, $E_{k\sigma \pm} \to E_{k-\sigma \pm}$), suggesting a $d$-wave AM occurs.  

For the Ising-Kondo lattice model with $t_{-}=t_{+}=0$, the ground state is normal antiferromagnet (NAF) at half-filling; however, when the electron filling deviates significantly from half-filling, the antiferromagnet is absent.\cite{PRBYang2019} Hence, when we take the case of $t_{-} \neq t_{+}$ into account, it is plausible to conclude that AM does not occur when the electron density significantly deviates from half-filling, which aligns with the phase diagrams in Fig.~\ref{fig:phase}. In Fig.~\ref{fig:Aktn0p1}, we plot the spectral function $A(k, \omega)$ for $t_{-}-t_{+}=-0.2$. In parts (a) and (b) with $\mu=0.22$ (half-filling) corresponding to spin up and spin down respectively, we observe that the spectral function exhibits spin splitting, which is consistent with the AM state. In contrast, for parts (c) and (d) with $\mu=1.1$ (far from half-filling), there is no spin splitting, and hence the absence of altermagnetic order is confirmed. These results highlight the critical role of electron filling in AM.

\begin{figure}[H]
	\begin{centering}
		\includegraphics[width=0.43\textwidth]{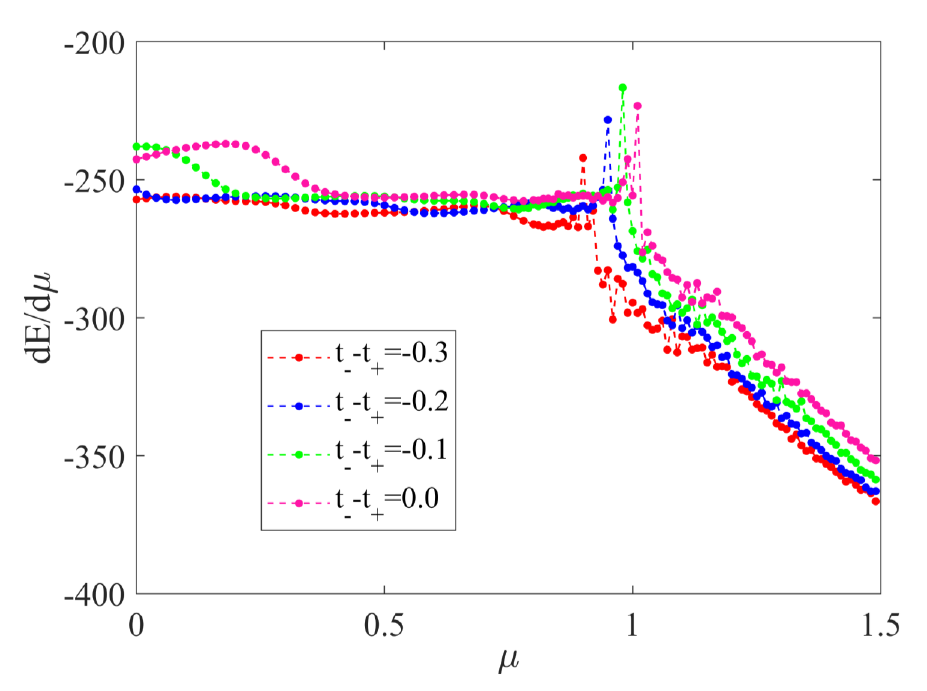}
		\par\end{centering}
	\protect\caption{
		\label{fig:dEdmu} The first derivative of energy $E$ with respect to the chemical potential $\mu$ in the Ising-Kondo lattice model, $dE/d\mu$, for different values of $t_{-}-t_{+}$. The other parameters are $J=3$ and $t_{+}=0.3$.}
\end{figure}

In order to determine the characteristics of the phase transition between AM/NAF and the spin density wave state (SDW) by tuning the chemical potential $\mu$ as shown in Fig.~\ref{fig:phase}, we plot $dE/d \mu$ in Fig.~\ref{fig:dEdmu}. This figure reveals peaks for all values of $t_{-}-t_{+}$. Therefore, the phase transition from AM/NAM to SDW turns out to be first order. (The details of the SDW phase can be found in the Appendix \ref{SDW}.)

\subsection{the J depending phase diagram at half-filling}\label{subsec3}

In the previous section, we examined the influence of NNNH and electron filling on the system's ground state, establishing that the AM state emerges only when $t_{-}-t_{+}\neq 0$ and the electron filling is close to half-filling. In this section, we will investigate the dependence of the AM state on the Ising-Kondo interaction $J$, and find that, at half-filling, the AM state remains stable as long as $J$ is larger than a critical value.

\begin{figure}[H]
	\begin{centering}
		\includegraphics[width=0.5\textwidth]{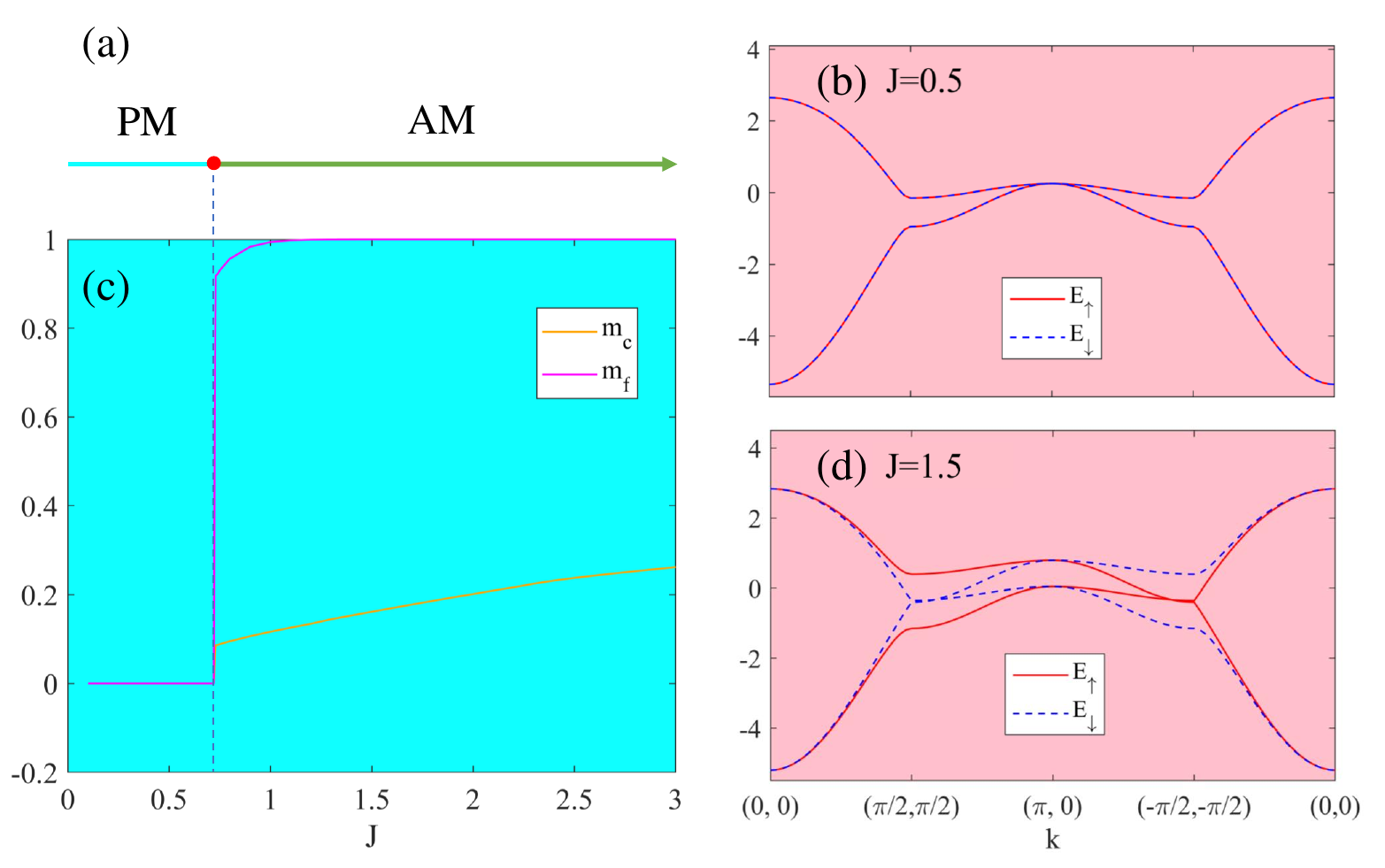}
		\par\end{centering}
	\protect\caption{(a) The schematic ground state phase diagram of the Ising-Kondo lattice model with alternating NNNH ($t_{-}-t_{+}=-0.2$) obtained using mean-field method. There exist two kinds of states, i.e., paramagnet (PM) and AM. (c) Mean-field order parameters $m_{c}$ and $m_{f}$ evolve with increasing Ising-Kondo coupling $J$ in the ground state of Ising-Kondo lattice model. (b) Spin-degenerated bands in the PM ($J=0.5$). (d) Spin-splitting bands in the AM ($J=1.5$). The $C_{4z}$ symmetry of bands indicates a d-wave AM. The other parameters are $t_{+}=0.3$ and $n_{c}=1$.
		\label{fig:meanfield}}
\end{figure}

First, we use mean-field theory to calculate the ground state of the system by varying Ising-Kondo interaction $J$. The antiferromagnetic mean-field Hamiltonian can be obtained by decoupling the Ising-Kondo coupling term as
\begin{eqnarray}
	\sum_{\sigma}\hat{S}^{z}_{j}\sigma \hat{c}^{\dagger}_{j\sigma} \hat{c}_{j\sigma} &\simeq& (-1)^{j+1}\frac{m_{f}}{2}\sum_{\sigma}\sigma \hat{c}^{\dagger}_{j\sigma} \hat{c}_{j\sigma} +(-1)^{j}m_{c}\hat{S}^{z}_{j} \nonumber \\
	&+& \frac{m_{f}m_{c}}{2}
\end{eqnarray}  
where the magnetic order parameters are defined by
\begin{eqnarray}
	&\langle \hat{S}^{z}_{j}\rangle=(-1)^{j+1}\frac{m_{f}}{2}\nonumber \\
	&\sum_{\sigma}\sigma \langle \hat{c}^{\dagger}_{j\sigma} \hat{c}_{j\sigma} \rangle=(-1)^{j}m_{c}. \nonumber
\end{eqnarray} 
Thus, the mean-field Hamiltonian reads as
\begin{eqnarray}
	\hat{H}&=&-\sum_{i,j,\sigma}t_{ij}\hat{c}_{i\sigma}^{\dagger}\hat{c}_{j\sigma}+\frac{Jm_{f}}{4}\sum_{j,\sigma}(-1)^{j+1}\sigma\hat{c}_{j\sigma}^{\dagger}\hat{c}_{j\sigma} \nonumber \\
	&+&\frac{J}{2}m_{c}\sum_{j}(-1)^{j}\hat{S}_{j}^{z}+\sum_{j}\frac{Jm_{f}m_{c}}{4}
\end{eqnarray}

After the Fourier transform $\hat{c}_{jA\sigma}=\frac{1}{\sqrt{N_{s}}}\sum_{k}e^{ik\cdot R_{j}}\hat{c}_{Ak\sigma}$ and $\hat{c}_{jB\sigma}=\frac{1}{\sqrt{N_{s}}}\sum_{k}e^{ik\cdot R_{j}}\hat{c}_{Bk\sigma}$, the Hamiltonian can be written as
\begin{eqnarray}
	\hat{H}(k)&=&\sum_{k,\sigma}[\varepsilon_{k}\hat{c}_{Ak\sigma}^{\dagger}\hat{c}_{Bk\sigma}+h.c+(\varepsilon_{k}^{AA}+\frac{Jm_{f}}{4}\sigma)\hat{c}_{Ak\sigma}^{\dagger}\hat{c}_{Ak\sigma} \nonumber \\
	&+&(\varepsilon_{k}^{BB}-\frac{Jm_{f}}{4}\sigma)\hat{c}_{Bk\sigma}^{\dagger}\hat{c}_{Bk\sigma}] \nonumber\\
	&+&\frac{J}{2}m_{c}\sum_{j}(-1)^{j}\hat{S}_{j}^{z}+\sum_{j}\frac{Jm_{f}m_{c}}{4} \nonumber \\
	&=&\hat{H}_{e}+\hat{H}_{s}+2N_{s}\frac{Jm_{f}m_{c}}{4}
\end{eqnarray} 
with $\hat{H}_{s}=\frac{J}{2}m_{c}\sum_{j}(-1)^{j}\hat{S}_{j}^{z}$ and 	
\begin{eqnarray}
	\hat{H}_{e}&=&\sum_{k,\sigma}[\varepsilon_{k}\hat{c}_{Ak\sigma}^{\dagger}\hat{c}_{Bk\sigma}+h.c  +(\varepsilon_{k}^{AA}+\frac{Jm_{f}}{4}\sigma)\hat{c}_{Ak\sigma}^{\dagger}\hat{c}_{Ak\sigma} \nonumber \\
	&+&(\varepsilon_{k}^{BB}-\frac{Jm_{f}}{4}\sigma)\hat{c}_{Bk\sigma}^{\dagger}\hat{c}_{Bk\sigma}]\nonumber \\
	&=&	\sum_{k,\sigma}\left(\begin{array}{cc}
		\hat{c}_{Ak\sigma}^{\dagger} & \hat{c}_{Bk\sigma}^{\dagger}\end{array}\right)\left(\begin{array}{cc}
		\varepsilon_{k\sigma}^{AA'} & \varepsilon_{k}\\
		\varepsilon_{k} & \varepsilon_{k\sigma}^{BB'}
	\end{array}\right)\left(\begin{array}{c}
		\hat{c}_{Ak\sigma}\\
		\hat{c}_{Bk\sigma}
	\end{array}\right) 
\end{eqnarray}
Here, $\varepsilon_{k\sigma}^{AA'}=\varepsilon_{k}^{AA}+\frac{Jm_{f}}{4}\sigma$, $\varepsilon_{k\sigma}^{BB'}=\varepsilon_{k}^{BB}-\frac{Jm_{f}}{4}\sigma$ and the summation over momentum $k$ is also performed within the reduced Brillouin zone. Therefore, its free energy can be expressed as
\begin{eqnarray}
	f&=&-\frac{1}{\beta}\sum_{k,\sigma}[\ln(1+e^{-\beta E_{k\sigma+}})+\ln(1+e^{-\beta E_{k\sigma-}})] \nonumber \\
	&-&\frac{2N_{s}}{\beta}\ln[2\cosh(\frac{\beta Jm_{c}}{4})]+2N_{s}\frac{Jm_{f}m_{c}}{4}
\end{eqnarray}
with $ E_{k\sigma \pm}=\frac{1}{2}[\pm\sqrt{(\varepsilon_{k}^{AA}-\varepsilon_{k}^{BB}+\frac{Jm_{f}}{2}\sigma)^{2}+4\varepsilon_{k}^{2}}+\varepsilon_{k}^{AA}+\varepsilon_{k}^{BB}-2\mu]$, where $\mu$ is considered to fix electron density. So the mean-field self-consistent equations can be derived by $\frac{\partial F}{\partial m_{c}}=0$ and $\frac{\partial F}{\partial m_{f}}=0$, i.e.,
\begin{eqnarray}
	m_{f}-\frac{\sinh(\frac{\beta Jm_{c}}{4})}{\cosh(\frac{\beta Jm_{c}}{4})}=0
\end{eqnarray}
\begin{eqnarray}
	&\sum_{k}&\frac{1}{N_{s}}[(f_{F}(E_{k\uparrow+})-f_{F}(E_{k\uparrow-})) \nonumber \\
	&\times& \frac{\varepsilon_{k}^{AA}-\varepsilon_{k}^{BB}+\frac{Jm_{f}}{2}}{\sqrt{(\varepsilon_{k}^{AA}-\varepsilon_{k}^{BB}+\frac{Jm_{f}}{2})^{2}+4\varepsilon_{k}^{2}}]}\nonumber \\
	&+&(f_{F}(E_{k\downarrow-})-f_{F}(E_{k\downarrow +})) \nonumber \\
	&\times&\frac{\varepsilon_{k}^{AA}-\varepsilon_{k}^{BB}-\frac{Jm_{f}}{2}}{\sqrt{(\varepsilon_{k}^{AA}-\varepsilon_{k}^{BB}-\frac{Jm_{f}}{2})+4\varepsilon_{k}^{2}}]}]+2m_{c}\nonumber\\
	&=&0.
\end{eqnarray}
where $f_{F}(x) = 1/(e^{x/T} + 1)$ is the standard Fermi distribution function.

By solving the above equations, we are able to determine all order parameters for different values of $J$ at half-filling, and the results of the order parameters and the corresponding $J$-dependent phase diagram are displayed in parts (c) and (a) of Fig.~\ref{fig:meanfield}, respectively. Observing Fig.~\ref{fig:meanfield}(c), when $J$ is larger than a critical value $J_{c}$ ($J_{c}\approx 0.72$), the system exhibits antiferromagnetic order. Otherwise, it is paramagnet (PM) phase. Furthermore, by checking spin-splitting quasiparticle energy bands, we can demonstrate that the system is AM when the local spin maintains antiferromagnetic alignment. For example, we calculate the dispersion for $J=1.5$ as shown in Fig.~\ref{fig:meanfield}(d), which reveals spin splitting indicative of AM. Additionally, the dispersion also displays $C_{4z}$ symmetry, implying a $d$-wave AM. For comparison, we also calculate the case of small $J$ (where $J=0.5$, shown in Fig.~\ref{fig:meanfield}(b)), where no spin splitting is observed in the bands, which is reasonable for the PM state.
\begin{figure}[H]
	\begin{centering}
		\includegraphics[width=0.4\textwidth]{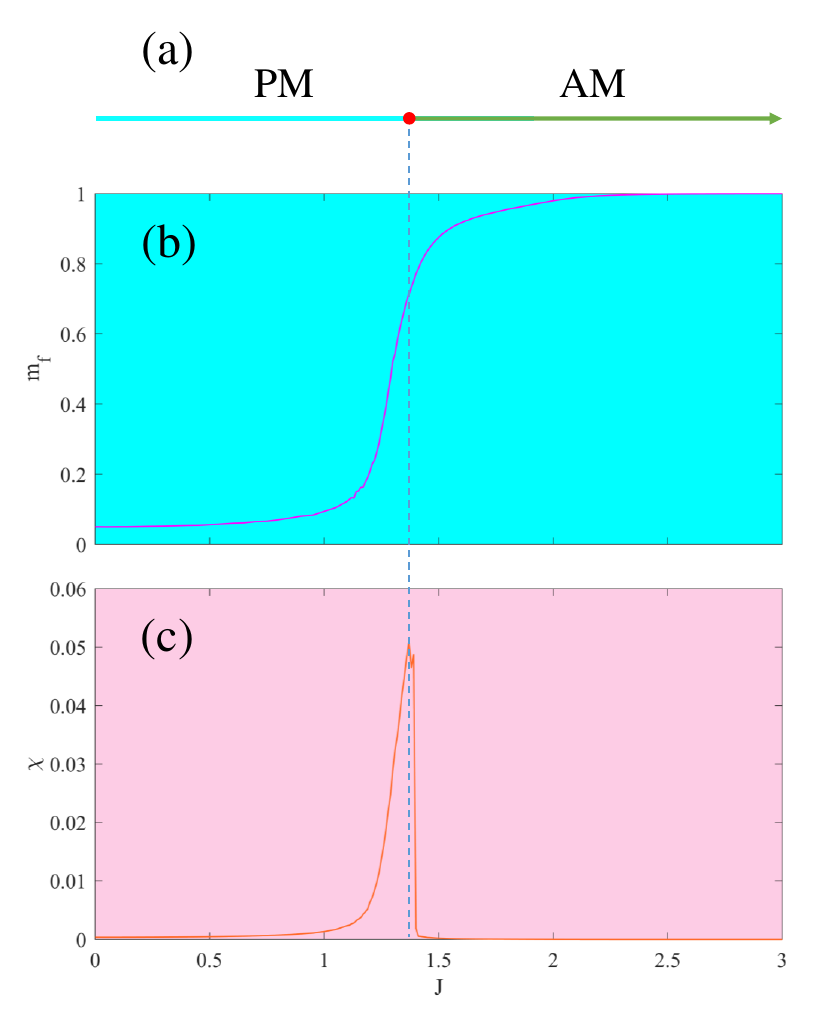}
		\par\end{centering}
	\protect\caption{(a) The schematic ground state phase diagram of the Ising-Kondo lattice model with alternating NNNH ($t_{-}-t_{+}=-0.2$) obtained through Monte Carlo simulations. There exist two kinds of states, i.e., PM and AM. (b) The order parameter $m_{f}$ evolves with increasing Ising-Kondo coupling $J$ in the ground state of the Ising-Kondo lattice model. (c) The susceptibility of the local spin $\chi$ as a function of Ising-Kondo coupling strength $J$ . The other parameters are $t_{+}=0.3$ and $n_{c}=1$.
		\label{fig:mf_MC}}
\end{figure}

\begin{figure}[H]
	\begin{centering}
		\includegraphics[width=0.43\textwidth]{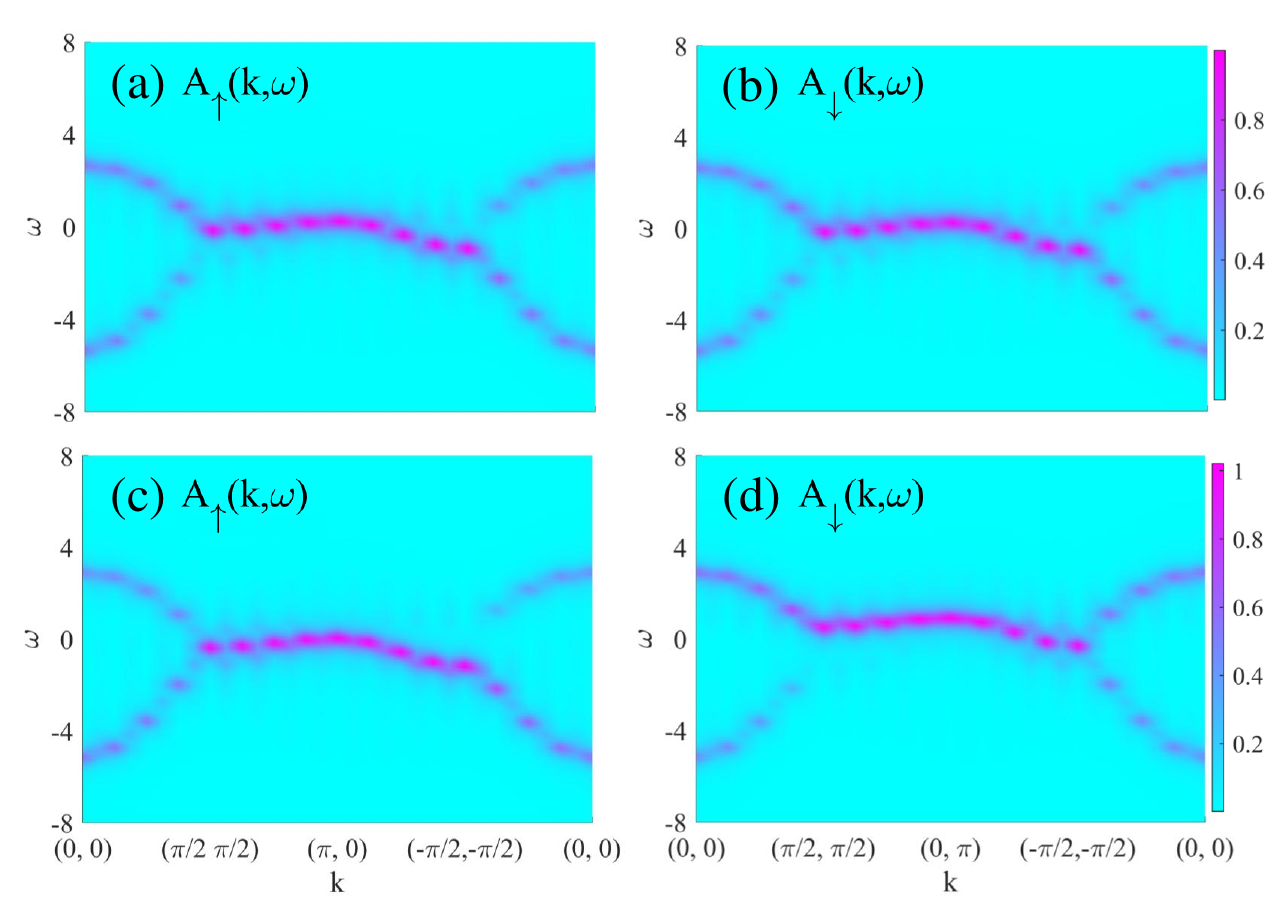}
		\par\end{centering}
	\protect\caption{The spectral function $A(k,\omega)$ for different Ising-Kondo interaction $J$ with $t_{-}-t_{+}=-0.2$ obtained through Monte Carlo simulations. (a) and (b) correspond to $J=0.7$ for spin up and spin down, respectively, in PM; (c) and (d) correspond to the case of $J=1.8$ in AM. The other parameters are $n_{c}=1$ and $t_{+}=0.3$.
		\label{fig:Ak_diff_J}}
\end{figure}

Then we exactly solve this model for different $J$ by LMC, obtaining the order parameter $m_{f}$ (Fig.~\ref{fig:mf_MC}(b)) and its corresponding phase diagram (Fig.~\ref{fig:mf_MC}(a)). In Fig.~\ref{fig:mf_MC}(b), the local spin changes from $0$ to $1$ with increasing $J$, suggesting a transition from PM to antiferromagnet, and therefore we can further conclude that the system undergoes a phase transition from PM to AM. This conclusion is supported by Fig.~\ref{fig:Ak_diff_J}, in which we observe spin splitting at large $J$, while no such splitting is present for small $J$. Clearly, these results qualitatively agree with those obtained from mean-field theory, although the value $J$ at the transition point, which is determined by the peak of the magnetic susceptibility (as shown in Fig.~\ref{fig:mf_MC}(c)), is larger than that predicted by mean-field theory. This discrepancy is reasonable because the mean-field method neglects the influence of fluctuation. Finally, we can find that in both cases, as long as $J>1.5$, the alternating NNNH induces AM at half-filling, making the choice of $J=3$ in section~\ref{subA} reasonable.

\subsection{the $(\delta, t_{1})$ phase diagram at half-filling}

\begin{figure}[H]
	\begin{centering}
		\includegraphics[width=0.41\textwidth]{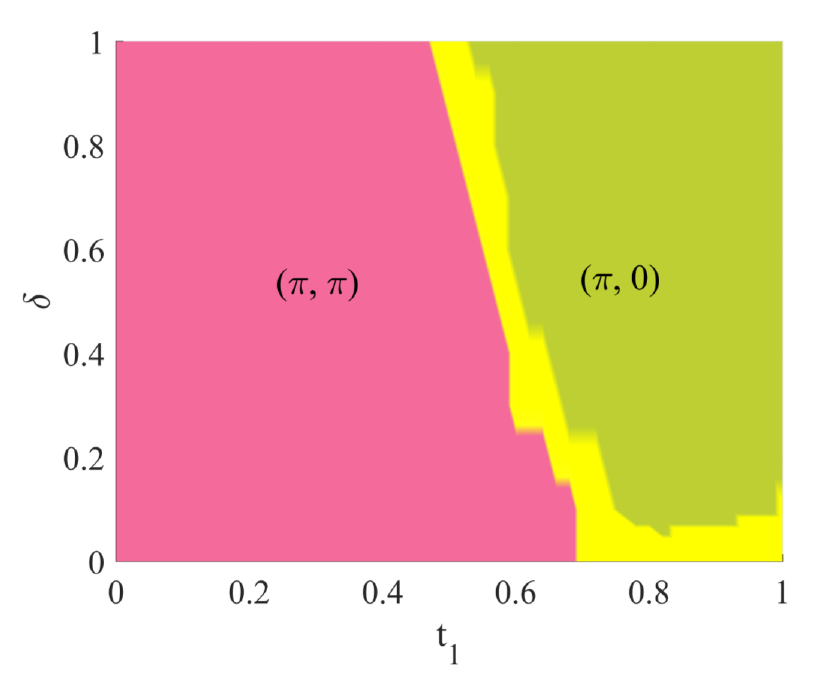}
		\par\end{centering}
	\protect\caption{
		\label{fig:phase3}The schematic ground state phase diagram of the Ising-Kondo lattice model vs $\delta$ and $t_{1}$. The other parameters are $J=5$ and $n_{c}=1$.}
\end{figure}

As described in Ref.~\onlinecite{Das2024} and also confirmed by previous sections, the staggered NNNH is one of the essential conditions for realizing AM. Therefore, in this section, we will discuss in detail how the NNNH affects the ground state of the Ising-Kondo lattice model at half-filling. To highlight the effect of NNNH, we introduce two new parameters, $t_{1}$ and $\delta$, to redefine NNNH, specifically $t_{+}=t_{1}(1+\delta)$ and $t_{-}=t_{1}(1-\delta)$. Moreover, we set $J=5$ in this section. 
\begin{figure}[H]
	\begin{centering}
		\includegraphics[width=0.43\textwidth]{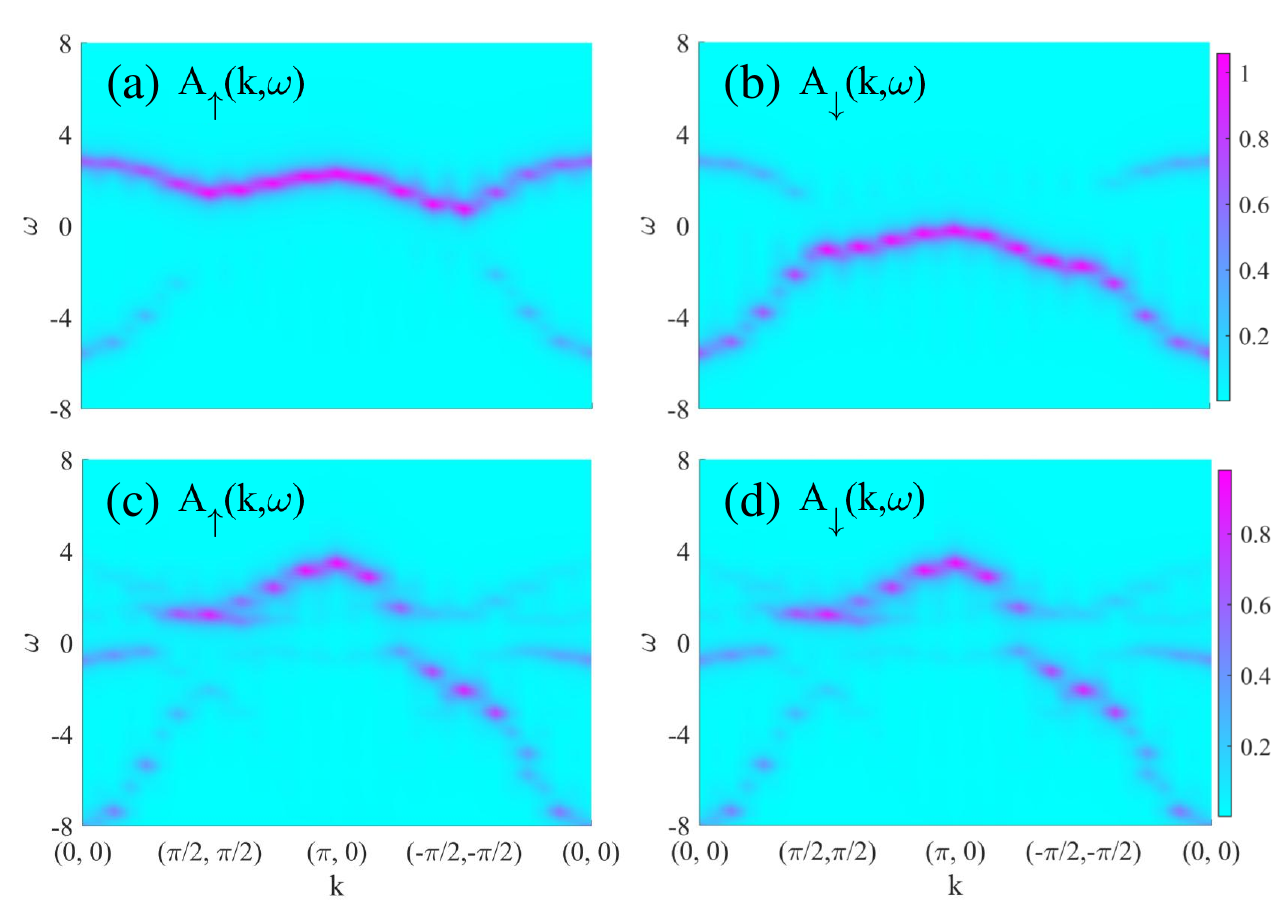}
		\par\end{centering}
	\protect\caption{
		\label{fig:AK_J5}The spectral function $A(k,\omega)$ for different NNNH parameters. (a) and (b) correspond to the case of $t_{1}=0.3$, representing to spin-up and spin-down states, respectively, in $(\pi, \pi)$ order; (c) and (d) correspond to the case of $t_{1}=0.9$ in $(\pi, 0)$ order. The other parameters are $J=5$, $\delta=0.3$ and $n_{c}=1$.}
\end{figure}

By tuning the parameters $t_{1}$ and $\delta$, we calculate the characteristic wave-vector distribution of the structure factor and obtain the $(\delta, t_{1})$ phase diagram (shown in Fig.~\ref{fig:phase3}). Three distinct orders are identified: $(\pi, \pi)$ order, $(\pi, 0)$ order, and disorder corresponding to the crossover from $(\pi, \pi)$ to $(\pi, 0)$. In the phase diagram, we observe that when $t_{1}$ is approximately smaller than $0.4$, the $(\pi, \pi)$ order is always stabilized. Obviously, when $t_{-}\neq t_{+}$ and the antiferromagnet of local spin order (i.e., $(\pi, \pi)$ order) is stable, the system will exhibit AM, as confirmed by the example in Fig.~\ref{fig:AK_J5} (a) and (b), where spin splitting is observed in the spectral function.

However, the $(\delta, t_{1})$ phase diagram also reveals the presence of striped $(\pi, 0)$ order. 
Here, we focus on this striped order $(\pi,0)$. In this case, the Hamiltonian can be written as 
\begin{eqnarray}
	\hat{H}=\hat{H}_{0}+\frac{J}{4}\sum_{j,\sigma}(-1)^{j_{x}}\sigma \hat{c}_{j\sigma}^{\dagger}\hat{c}_{j\sigma}
\end{eqnarray}

By performing a Fourier transform on the operators associated with the A and B sublattices, we can rewrite the Hamiltonian in momentum space:
 \begin{eqnarray}
 	\hat{H}(k)&=&\sum_{k,\sigma}\big[\varepsilon_{k}\hat{c}_{Ak\sigma}^{\dagger}\hat{c}_{Bk\sigma}+h.c.\nonumber \\
        &+&(\varepsilon_{k}^{AA}-\mu)\hat{c}^{\dagger}_{Ak\sigma}\hat{c}_{Ak\sigma}+(\varepsilon_{k}^{BB}-\mu)\hat{c}^{\dagger}_{Bk\sigma}\hat{c}_{Bk\sigma}\nonumber \\
        &+&\frac{J}{4}\sigma(\hat{c}^{\dagger}_{Ak\sigma}\hat{c}_{A(k+Q)\sigma}+\hat{c}^{\dagger}_{Bk\sigma}\hat{c}_{B(k+Q)\sigma})\big]
 \end{eqnarray}
with $Q=(\pi,0)$ as the characteristic wave vector of the striped order. Then, $\hat{H}(k)$ can be further written in the following form
\begin{eqnarray}
	\hat{H}(k)&=&\sum_{k,\sigma}\Big[\frac{1}{2}\varepsilon_{k}\hat{c}^{\dagger}_{Ak\sigma}\hat{c}_{Bk\sigma}+h.c. \nonumber \\ 
	&+&\frac{1}{2}\varepsilon_{k+Q}\hat{c}^{\dagger}_{A(k+Q)\sigma}\hat{c}_{B(k+Q)\sigma}+h.c. \nonumber \\
	&+&\frac{1}{2}(\varepsilon_{k}^{AA}-\mu)\hat{c}_{Ak\sigma}^{\dagger}\hat{c}_{Ak\sigma} +\frac{1}{2}(\varepsilon_{k}^{BB}-\mu)\hat{c}_{Bk\sigma}^{\dagger}\hat{c}_{Bk\sigma}\nonumber \\
	&+&\frac{1}{2}(\varepsilon_{k+Q}^{AA}-\mu)\hat{c}_{A(k+Q)\sigma}^{\dagger}\hat{c}_{A(k+Q)\sigma}	\nonumber \\
	&+&\frac{1}{2}(\varepsilon_{k+Q}^{BB}-\mu)\hat{c}_{B(k+Q)\sigma}^{\dagger}\hat{c}_{B(k+Q)\sigma}\nonumber \\
	&+&\frac{J}{8}\sigma(\hat{c}_{Ak\sigma}^{\dagger}\hat{c}_{A(k+Q)\sigma}+\hat{c}_{Bk\sigma}^{\dagger}\hat{c}_{B(k+Q)\sigma})+h.c.\Big] \nonumber \\
\end{eqnarray}

Introducing the four-component spinor 
\begin{eqnarray}
	\hat{\psi}_{k\sigma}^{\dagger}=(\hat{c}^{\dagger}_{Ak\sigma},\hat{c}^{\dagger}_{Bk\sigma},\hat{c}^{\dagger}_{A(k+Q)\sigma},\hat{c}^{\dagger}_{B(k+Q)\sigma}),
\end{eqnarray}
the Hamiltonian reads
\begin{eqnarray}
	\hat{H}(k)=\sum_{k,\sigma}\hat{\psi}_{k\sigma}^{\dagger}H_{\sigma}(k)\hat{\psi}_{k\sigma}
\end{eqnarray}
where
\begin{eqnarray}
	H_{\sigma}(k)=\left(\begin{array}{cccc}
		\frac{\varepsilon_{k}^{AA}-\mu}{2} & \frac{\varepsilon_{k}}{2} & \frac{J}{8}\sigma & 0\\
		\frac{\varepsilon_{k}}{2} & \frac{\varepsilon_{k}^{BB}-\mu}{2}& 0 &\frac{J}{8}\sigma\\
		\frac{J}{8}\sigma& 0 &\frac{\varepsilon_{k+Q}^{AA}-\mu}{2}&\frac{\varepsilon_{k+Q}}{2} \\
		0 & \frac{J}{8}\sigma& \frac{\varepsilon_{k+Q}}{2}&\frac{\varepsilon_{k+Q}^{BB}-\mu}{2}
	\end{array}\right)\nonumber
\end{eqnarray}
and this form of $H_{\sigma}(k)$ is independent of the spin index $\sigma$. Consequently, the energy bands of $H(k)$ remain spin-degenerated. This conclusion is fully consistent with the numerical results presented in Fig.~\ref{fig:AK_J5} (c) and (d), where no spin splitting is observed in the spectral function. Therefore, although the alternating NNNH can generate both $(\pi,0)$ and $(\pi,\pi)$ order of local spins in the Ising-Kondo lattice model, the AM phase emerges exclusively in association with $(\pi, \pi)$ order.

\section{Discussion}\label{sec3}

\subsection{impurity} \label{sec3C}
Several physical effects caused by impurities in altermagnets have already been studied. The Kondo effect induced by a single magnetic impurity has been investigated by Diniz and Vernek, who predicted the suppression of the Kondo temperature in altermagnets.\cite{PRBDiniz2024} The well-known RKKY interaction between two spinful impurities in altermagnets was recently explored in Refs. \onlinecite{ArxivLee2023,PRBAmundsen2024}. The RKKY interaction exhibits the characteristic behavior of $d$-wave altermagnets, specifically a $C_{4z}$-symmetric oscillating pattern characterized by multiple periods and angular dependence, resulting in anisotropy. At the same time, in Ref. \onlinecite{ArxivLee2023}, the single magnetic impurity problem was also addressed, suggesting that the Kondo temperature may be enhanced or reduced, depending on the band structure and the electron density, which is inconsistent with the conclusions drawn by Diniz and Vernek. In addition, researchers have discussed impurity-induced Friedel oscillations of the local density of states in altermagnets, noting that the period of these oscillations exhibits strongly anisotropic and direction-dependent behavior. \cite{PRBChen2024,PRBSukhachov2024} However, the aforementioned studies begin with a continuum model in momentum space, while we believe that exploring the impurity problem in real space will yield significant insights.

\begin{figure}[H]
	\begin{centering}
		\includegraphics[width=0.25\textwidth]{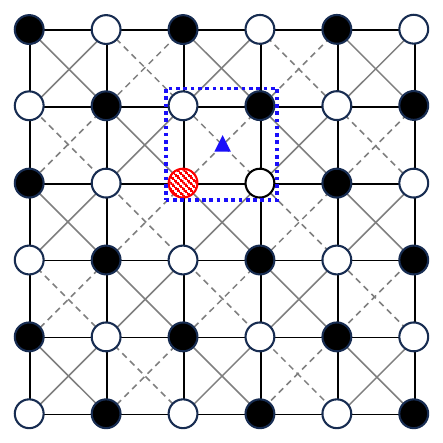}
		\par\end{centering}
	\protect\caption{
		\label{fig:impurity} The impurity (blue triangle) configuration used in the calculation for AM state.}
\end{figure}

Now, we consider the effect of a single non-magnetic impurity, which is assumed to be located at the center of a plaquette (the location of the blue triangle in Fig.~\ref{fig:impurity}), and which only affects the hopping energy of electron passing through it. Thus, we have the following impurity Hamiltonian:
\begin{eqnarray}
	\hat{H}_{imp} &=& V\sum_{\sigma}(\hat{c}_{(j_{x},j_{y})\sigma}^{\dagger} \hat{c}_{(j_{x+1},j_{y+1})\sigma} \nonumber \\
	&+&\hat{c}_{(j_{x+1},j_{y})\sigma}^{\dagger} \hat{c}_{(j_{x},j_{y+1})\sigma}+h.c)\nonumber
\end{eqnarray}
Here, $V$ represents the strength of the impurity effect, and the site $j=(j_{x},j_{y})$, i.e., the red site in Fig.~\ref{fig:impurity}, conveniently describes the location of impurity. 

\begin{figure}[H]
	\begin{centering}
		\includegraphics[width=0.41\textwidth]{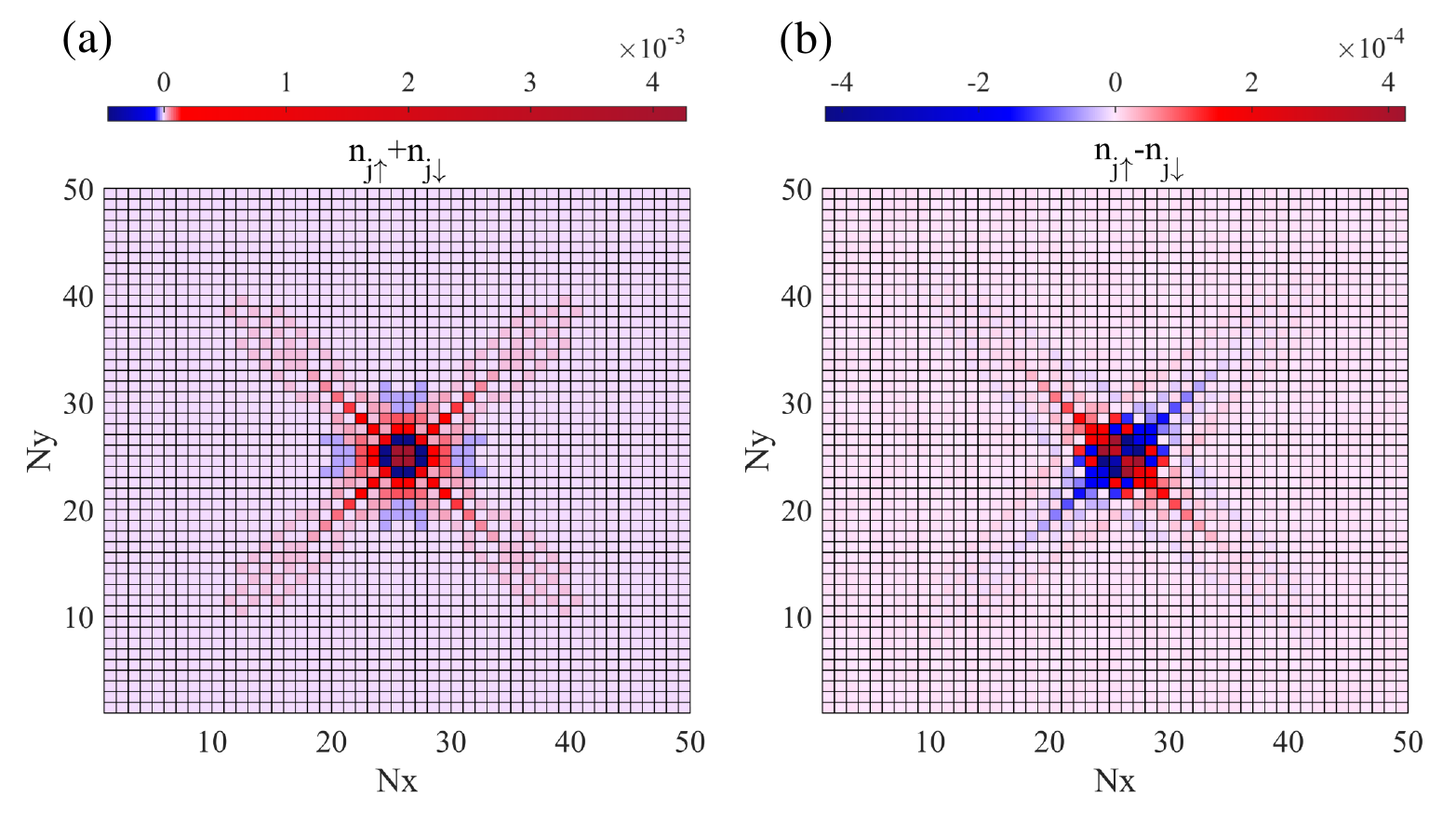}
		\par\end{centering}
	\protect\caption{
		\label{fig:impurity_nj} (a)The particle distribution $n_{j+}=n_{j \uparrow} + n_{j \downarrow}$ and (b) the magnetization distribution $n_{j-}=n_{j \uparrow} - n_{j \downarrow}$ in real space. Here, $J=2$, $t_{-}-t_{+}=-0.2$, $t_{+}=0.3$, $V=0.03$ and $n_{c}=1$.}
\end{figure}

We consider a $50 \times 50$ lattice with $j=(25,25)$, where the local spin configuration is antiferromagnetic, indicating the system is in the AM state. To illustrate the effect of impurity, we calculate the particle distribution $n_{j+}=n_{j \uparrow} + n_{j \downarrow}$ and the magnetization distribution $n_{j-}=n_{j \uparrow} - n_{j \downarrow}$ in real space, as shown in Fig.~\ref{fig:impurity_nj}. Fig.~\ref{fig:impurity_nj}(a) displays the particle distribution $n_{j+}$, in which the $C_{4z}$ symmetry is clearly observed. In contrast, the magnetization distribution $n_{j-}$ shown in Fig.~\ref{fig:impurity_nj}(b) exhibits broken symmetry, with polarity appearing between spin-up and spin-down states, which is consistent with the expected behavior for an AM state with $d$-wave symmetry. Therefore, these characteristics suggest that the system exhibits $d$-wave AM, as already supported by Fig.~\ref{fig:disp}(a), where the bands display $C_{4z}$ symmetry in momentum space.

\subsection{some observable} \label{sec3A}

Because of the spin-splitting bands, a spin-resolved conductivity is expected in the AM phase. Therefore, we will calculate the conductivity to validate this expectation. According to linear-response theory, the zero-temperature conductivity has the following expression (with a detailed derivation in the Appendix \ref{StCon})
\begin{eqnarray}
	\sigma_{\alpha \alpha}^{\sigma}=\frac{e^2}{\hbar}\frac{\pi}{N_{s}}\sum_{k}\mathrm{Tr}[\partial_{k_{\alpha}}H_{\sigma}(k)A_{\sigma}(k)\partial_{k_{\alpha}}H_{\sigma}(k)A_{\sigma}(k)].\nonumber
\end{eqnarray}
Here, the zero-frequency spectral function is defined as $A_{\sigma}(k)=-\frac{1}{\pi}\mathrm{Im} G_{\sigma}^{R}(k,\omega=0)$. In the AM state with $J=2$, for the sake of achieving a finite conductivity, and $t_{-}-t_{+}=-0.2$, we find $\sigma_{\alpha \alpha}^{\uparrow}=6.4581 \frac{e^{2}}{\hbar}$, $\sigma_{\alpha \alpha}^{\downarrow}=0.468 \frac{e^{2}}{\hbar}$, where the direction of conductivity is given by $(\alpha, \alpha)=(1,1)$. (A damping factor $\Lambda=0.02$ is used to obtain finite comductivity and the system size is $1000\times 1000$.) Inequality $\sigma_{\alpha \alpha}^{\uparrow}\neq \sigma_{\alpha \alpha}^{\downarrow}$ implies that there is indeed a nonzero spin-resolved current, which is essential for spintronics \cite{Bai2022}, and provides the feasibility for spintonics applications in heavy fermion compounds with AM-like states

\begin{figure}[H]
	\begin{centering}
		\includegraphics[width=0.42\textwidth]{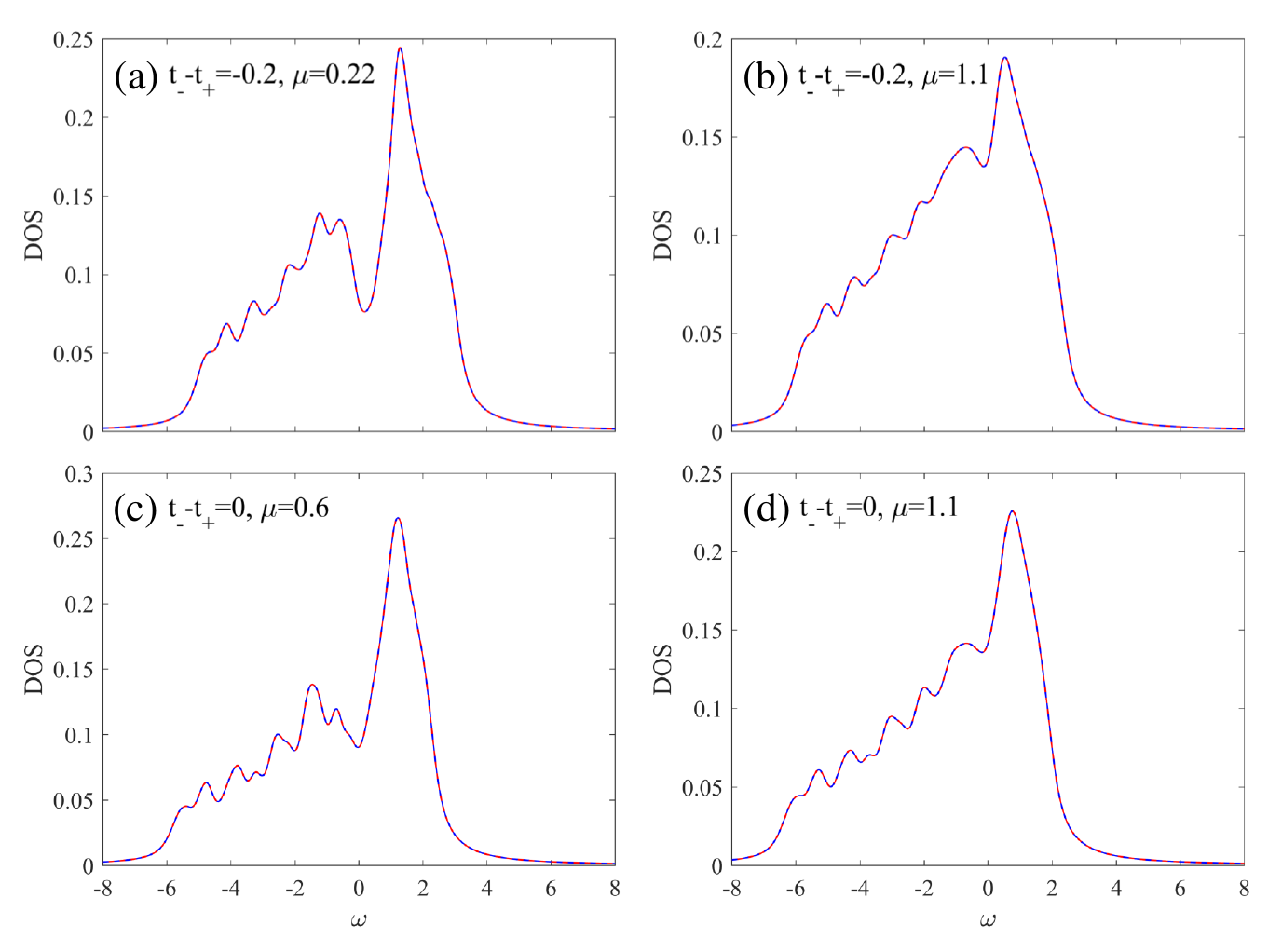}
		\par\end{centering}
	\protect\caption{
		\label{fig:DOS} The density of states $DOS$ for different NNNH parameters and electronic filling: (a) $t_{-}-t_{+}=-0.2$, $\mu=0.22$ (half-filling), corresponding to AM; (b) $t_{-}-t_{+}=-0.2$, $\mu=1.1$ (far away half-filling), corresponding to SDW; (c) $t_{-}-t_{+}=0$, $\mu=0.6$ (half-filling), corresponding to NAF; (d) $t_{-}-t_{+}=0$, $\mu=1.1$ (far away half-filling), corresponding to SDW. There is no spin splitting for all states. The other parameters are $t_{+}=0.3$ and $J=3$. Here, the red solid curve represents spin-up, and the blue dashed curve does spin-down.}
\end{figure}

Finally, we plot the density of states for the AM (Fig.~\ref{fig:DOS}(a)), NAF (Fig.~\ref{fig:DOS}(c)) and SDW (Fig.~\ref{fig:DOS}(b) and Fig.~\ref{fig:DOS}(d)) with $J=3$. It is clear that, for all cases, the density of states for spin-up and spin-down does not split. For the NAF and SDW states, the absence of spin splitting is evident due to the completely overlapping spectral functions for spin up and spin down. But, for the AM state, we also observe that there is no spin splitting. This is because, in the AM state, both the band structure and the spectral function possess the symmetry of a $\frac{\pi}{2}$ rotation along with a spin flip, as confirmed by Fig.~\ref{fig:disp}(a). And when calculating the DOS, we need sum the entire momentum space. From Fig.~\ref{fig:disp} (a) and (b), an energy gap can be observed, confirming that the system is an insulator. However, in the Fig.~\ref{fig:DOS} (a) and (c), we also find a finite DOS, which is mainly due to the finite dissipation factor $\Gamma=0.3$.

\subsection{Finite temperature}\label{sec3B}
\begin{figure}[H]
	\begin{centering}
		\includegraphics[width=0.41\textwidth]{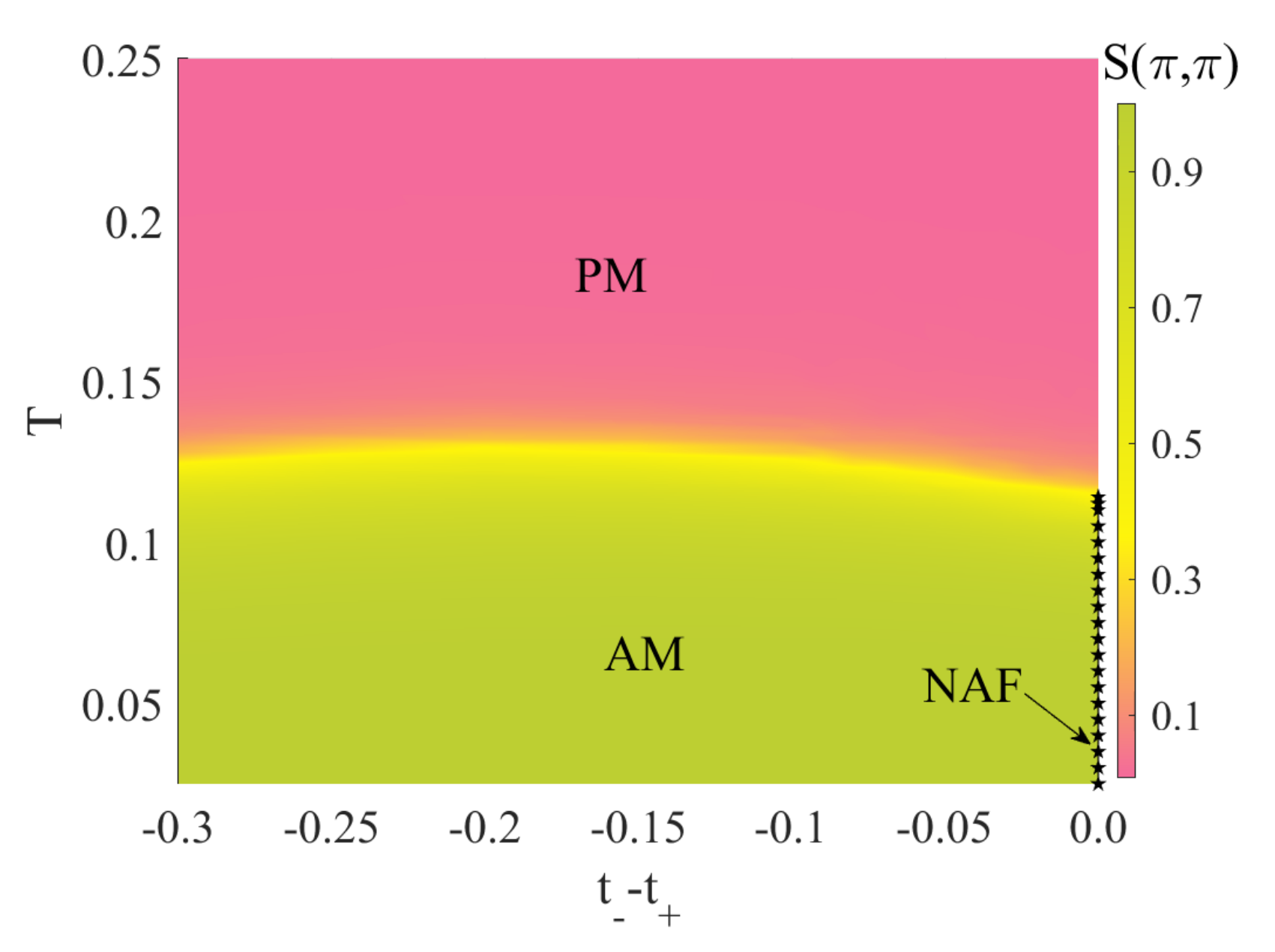}
		\par\end{centering}
	\protect\caption{The schematic finite-temperature phase diagram of the Ising-Kondo lattice model as a function of  NNNH ($t_{-}-t_{+}$) and temperature $T$. There exist three types of states, namely AM, NAF (corresponding to $t_{-}-t_{+}=0$, i.e., the solid line with pentagram) and PM. The other parameters are $J=3$, $t_{+}=0.3$ and $n_{c}=1$.
		\label{fig:phaseFTS} }
\end{figure}

\begin{figure}[H]
	\begin{centering}
		\includegraphics[width=0.41\textwidth]{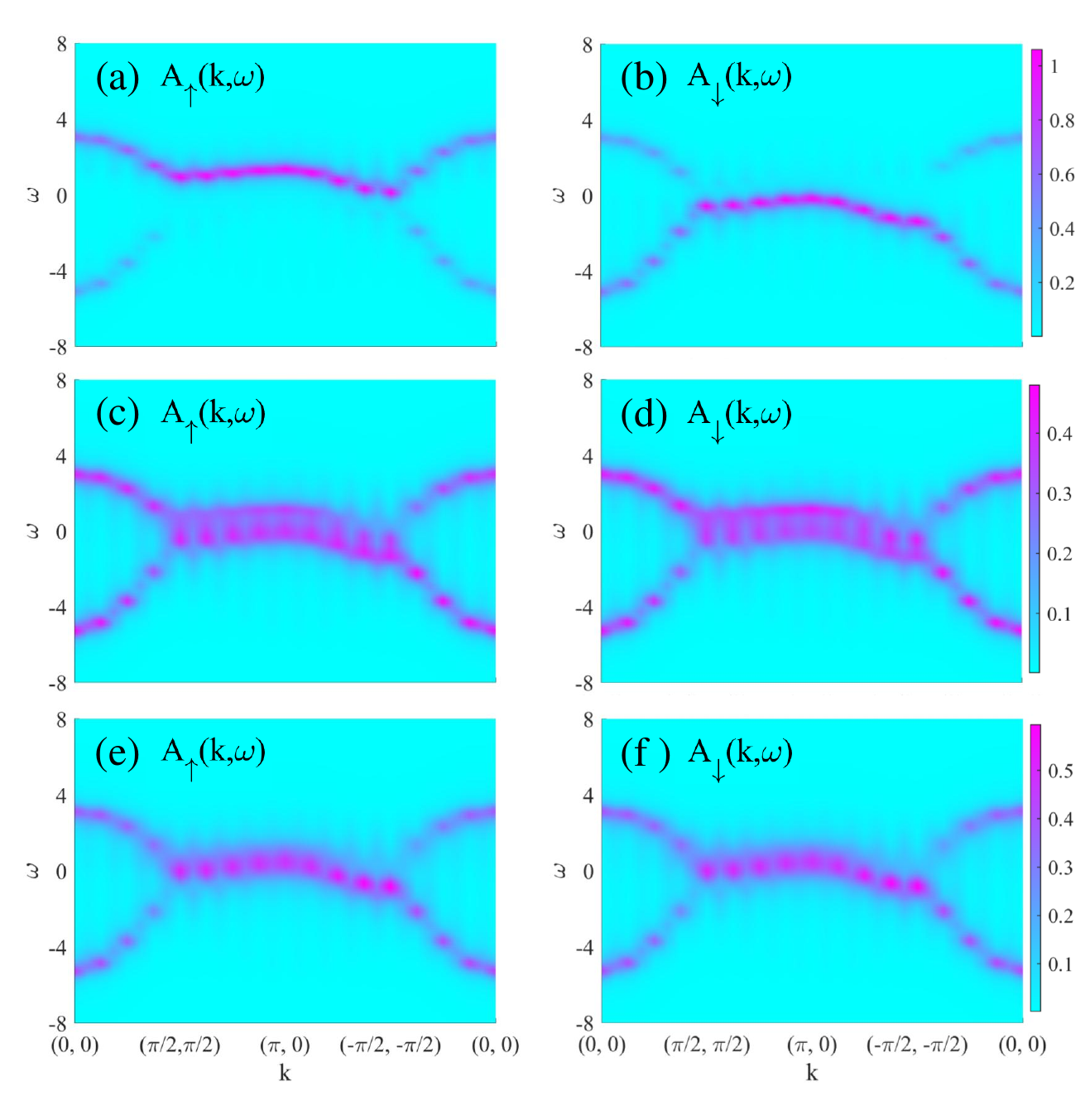}
		\par\end{centering}
	\protect\caption{
		\label{fig:AkFT} The spectral function $A(k,\omega)$ for different temperature $T$ with  $t_{-}-t_{+}=-0.2$. (a) and (b) correspond to the ground state($T=0.04$), i.e., AM; (c) and (d) correspond to the intermediate temperature case ($T=0.13$); (e) and (f) correspond to the high temperature case ($T=0.4$). The left panels represent spin up, and the right panels do spin down. The other parameters are $n_{c}=1$, $t_{+}=0.3$ and $J=3$.}
\end{figure}

At finite temperature $T$, one must sum all the configurations of the effective Ising spin $\{q_{j}\}$, which can only be performed via Monte Carlo simulation. We consider periodic $N=L\times L$ lattices with $L$ up to $16$. The resulting phase diagram is shown in Fig.~\ref{fig:phaseFTS}. Here, when $t_{-}-t_{+}\neq 0$, the AM is stable at low $T$, whereas for $t_{-}-t_{+}=0$, there is NAF at low $T$. With increasing $T$, there is a thermodynamic transition from AM or NAF to PM. Similarly to the case of the ground state, the phase diagram is determined by the structure factor $S(\pi, \pi)$.  According to the discussion above, we know that at low $T$, the system is either in the AM state or in the NAM state, so we take $(\pi, \pi)$ as the characteristic wave-vector, and the value of the structure factor $S(\pi, \pi)$ is $1$. At high $T$, $S(\pi, \pi)$ approaches zero and signals a transition to PM phase. For the sake of clearly displaying the characteristics of the AM, we also calculate the spectral function at low temperature (that is, ground state), intermediate temperature and high temperature, as shown in Fig.~\ref{fig:AkFT}. Obviously, only at low temperature does spin splitting exist, because thermal fluctuations at high $T$ will destroy the AM state.

\section{Conclusion}\label{sec4}
 In this paper, we explore AM-like phases in the Ising-Kondo lattice model, a prototypical model of heavy fermion systems, on a square lattice with alternating NNNH using LMC simulations. We observe the characteristics of $d$-wave AM states, including spin-splitting bands, spectral function and spin-resolved conductivity. Using these features, we determine the parameter region of $n_c$, $J$ and NNNH where the AM phase occurs in the ground state, finding that AM states can be robust across a broad range of the Kondo interaction strengths, doping levels and alternating NNNH. Additionally, we construct the phase diagram with respect to temperature, and, as expected, thermal fluctuation disrupts the long-range order, causing the system to undergo a phase transition from AM to PM. Finally, we examine the impact of non-magnetic impurity in real space, employing the particle and magnetization distribution analyses to reaffirm the $d$-wave symmetry of the AM phase.
 
 As is well known, alternating NNNH is one of effects arising from non-magnetic atoms, but according to the Ref.~\onlinecite{Brekke2023}, the electrons of non-magnetic atoms also influence the interaction between the local spins. Hence, we will take into account other effects resulting from electrons of non-magnetic atoms, which can describe more realistic materials and provide a pathway for searching for heavy-fermion compounds.

\section*{Acknowledgments}
This work was supported by the Supercomputing Center of Lanzhou University, which provided essential computational resources. We also acknowledge support from the National Natural Science Foundation of China (Grants No. 12247101), the Fundamental Research Funds for the Central Universities (Grants No. lzujbky-2024-jdzx06), the Natural Science Foundation of Gansu Province (Grants No. 22JR5RA389 and No. 25JRRA799), and the "111 Center" under Grant No. B20063.

\appendix



\section{$(t_{-}-t_{+}, n_{c})$ phase diagram on $4\times 4$ lattice} \label{SDW}
\begin{figure}[H]
	\begin{centering}
		\includegraphics[width=0.4\textwidth]{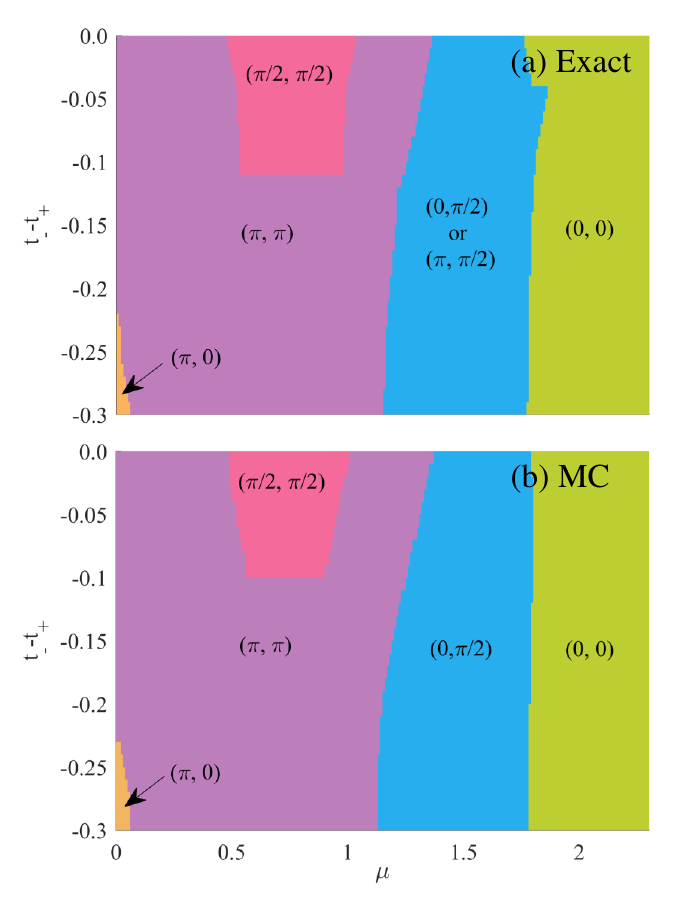}
		\par\end{centering}
	\protect\caption{
		\label{fig:struc4T4}The schematic ground state phase diagram of the Ising-Kondo lattice model with NNNH as a function of $t_{-}-t_{+}$ and $\mu$ is obtained by two methods: (a) exact calculation ($T=0.001$), and (b) MC simulations ($T=0.04$). The other parameters are $J=3$ and $t_{+}=0.3$.}
\end{figure}

As we have seen in the phase diagram Fig.~\ref{fig:phase}, when $n_c$ significantly deviates from half-filling, SDW is referred. This phase is not easy to be obtained by LMC on a finite but not too small size lattice for many reasons. Hence, in order to explain this case and prove that the state with a large electron density is indeed SDW, we calculate the structure factor on a $4\times4$ lattice by exact calculation and LMC. In the exact calculation, we sweep all spin configurations of $4\times4$ lattice, pick out the ground states, and calculate dominating wave-vector via the spin structure factor which is shown in Fig.~\ref{fig:struc4T4} (a). The results obtained by LMC are shown in Fig.~\ref{fig:struc4T4} (b). Both figures tell us that there exists characteristic wave-vector in the ground state, and it is reasonable to believe that this phase is SDW for large $\mu$ in Fig.~\ref{fig:phase}.

In addition, even though Fig.~\ref{fig:struc4T4} (b) qualitatively agrees with Fig~\ref{fig:struc4T4} (a), there are some differences between them, especially in the intermediate $\mu$. This is because in the range of intermediate $\mu$, there are many configurations with energies that are very close to each other (for example, in the case of $t_{-}-t_{+}=-0.06$ and $\mu=1.7$, $E_{1}-E_{0}=0.0389$, $E_{2}-E_{0}=0.0562$, where $E_{0}$, $E_{1}$ and $E_{2}$ are the lowest, second lowest, and third lowest energies, respectively), and the probability of state selected in LMC is determined by its energy. Thus, the probabilities of these states are almost identical, and the difference between the exact calculation and LMC occurs when the $\mu$ is at the intermediate value or the boundary of the phase transition. Furthermore, in contrast to Fig.~\ref{fig:phase}, when the electron density is near half-filling, there are two other phases, $(\pi,0)$ and $(\pi/2,\pi/2)$, which result from the finite-size lattice combined with periodic boundary condition. It is precisely for these reasons that make it difficult for us to clearly determine the SDW state on the $16\times16$ lattice by LMC. But the good news is that when the system is close to half-filling, the energy of $(\pi, \pi)$ order is much lower than that of other states, and hence the AM can be well identified in our calculation.

\section{Static conductivity }\label{StCon}

The conductivity tensor is a $2\times 2$ matrix for each spin
\begin{equation}
	\sigma^{\sigma} = {\left[ \begin{array}{ccc}
		\sigma_{\tilde{x} \tilde{x}}^{\sigma} & \sigma_{\tilde{x} \tilde{y}}^{\sigma} \\
	    \sigma_{\tilde{y} \tilde{x}}^{\sigma} & \sigma_{\tilde{y} \tilde{y}}^{\sigma} 
    \end{array}\right]}, \nonumber
\end{equation}
and the DC conductivity $\sigma_{\alpha \beta}^{\sigma}(\omega \rightarrow 0)$ with a specific spin $\sigma$ can be determined by the following formalism:
\begin{eqnarray}
	\sigma_{\alpha \beta}^{\sigma}=\lim_{\omega \rightarrow 0}\frac{\mathrm{Im}\Lambda_{\alpha \beta}^{\sigma}(\omega)}{\omega}. \nonumber
\end{eqnarray}  
where $\Lambda_{\alpha \beta}^{\sigma}(\omega)$ is the current-current correlation function. Hence, in order to calculate conductivity along $(\alpha, \alpha)$ direction,  we first have to derive an explicit expression for this current-current correlation function $\Lambda_{\alpha \alpha}^{\sigma}(\omega)$.  

As mentioned in the main text, the Hamiltonian in momentum $k$ space within the AM phase can be written as
\begin{eqnarray}\label{H_gkA}
	\hat{H}_{\sigma}(k)&=& \sum_{k} \left( \begin{array}{lr}\hat{c}_{Ak \sigma}^{\dagger} &\hat{c}_{Bk\sigma}^{\dagger}  \end{array}\right) \left( \begin{array}{cc} \varepsilon_{k\sigma}^{AA} &\varepsilon_{k} \\ \varepsilon_{k}&\varepsilon_{k\sigma}^{BB} \end{array}\right) \left( \begin{array}{l}\hat{c}_{Ak \sigma} \\ \hat{c}_{Bk\sigma}  \end{array}\right) \nonumber 
\end{eqnarray} 
where $\varepsilon_{k\sigma}^{AA}=\varepsilon_{k}^{AA}+\frac{J\sigma}{4}$, $\varepsilon_{k\sigma}^{BB}=\varepsilon_{k}^{BB}-\frac{J\sigma}{4}$, $\varepsilon_{k}^{AA} = -2t_{-} \cos(k_x+k_y) - 2t_{+} \cos(k_x-k_y)$, $\varepsilon_{k}^{BB}=-2t_{+}\cos(k_x+k_y)-2t_{-}\cos(k_x-k_y)$ and $\varepsilon_{k}=-2t[\cos(k_x)+\cos(k_y)]$. 
In this form of the Hamiltonian, the current-density operator is given by
\begin{eqnarray}
	\hat{J}_{e}^{\alpha\sigma}(q)=-e \sum_{k,a,b}\hat{c}^{\dagger}_{k,a,\sigma} \partial_{k_{\alpha}}H_{\sigma}^{ab}(k+q/2)\hat{c}_{k+q,b,\sigma} \nonumber
\end{eqnarray}
and $\partial_{k_{\alpha}} H_{\sigma}^{ab}(k+q/2)$ denotes the generalized velocity operator. Then the imaginary-time current-current correlation function can be written as 
\begin{eqnarray}
	\Lambda_{\alpha \alpha}^{\sigma}=\frac{1}{V}\langle \hat{T}_{\tau} \hat{J}^{\alpha\sigma}_{e}(q,\tau) \hat{J}^{\alpha\sigma}_{e}(-q,0)\rangle \nonumber
\end{eqnarray}
where $V$ is the volume of system. After performing a Fourier transform, the imaginary-frequency correlation function can be obtained
\begin{eqnarray}
	\Lambda_{\alpha \alpha}^{\sigma}=\frac{1}{V}\int d\tau e^{i\Omega_{n} \tau}\langle \hat{T}_{\tau} \hat{J}^{\alpha}_{e}(q,\tau)  \hat{J}^{\alpha}_{e}(-q,0)\rangle \nonumber
\end{eqnarray}
Only taking the limit $q \rightarrow 0 $ into the account, we can obtain the specific expression for the current-current correlation function
\begin{eqnarray}
	\Lambda_{\alpha \alpha}^{\sigma}(\Omega_{n}) &=& \frac{e^{2}}{V}\sum_{k,a,b}\sum_{k',a',b'}\int d\tau e^{i\Omega_{n} \tau} \partial_{k_{\alpha}} H_{\sigma}^{ab}(k) \nonumber \\
	&\times& \partial_{k_{\alpha}'} H_{\sigma}^{a'b'}(k') \langle \hat{T}_{\tau}\hat{c}^{\dagger}_{k a \sigma}(\tau)\hat{c}_{kb\sigma}(\tau) c^{\dagger}_{k' a' \sigma}\hat{c}_{k'b'\sigma} \rangle. \nonumber 
\end{eqnarray}

By using the Wick theorem, we can simplify the above equation as follows

\begin{eqnarray}
	\Lambda_{\alpha \alpha}^{\sigma}(\Omega_{n})&=&-\frac{e^{2}}{V} \sum_{k,a,b}\sum_{k',a',b'}\int d\tau e^{i\Omega_{n}\tau}\partial_{k_{\alpha}} H_{\sigma}^{ab}(k) \nonumber \\
	&\times& \partial_{k_{\alpha}'}H_{\sigma}^{a'b'}(k') \langle \hat{T}_{\tau} \hat{c}_{k' b'\sigma}\hat{c}_{k a\sigma}^{\dagger}(\tau)\rangle \langle\hat{T}_{\tau}\hat{c}_{k b \sigma}(\tau)\hat{c}^{\dagger}_{k' a' \sigma} \rangle\nonumber \\	    
	&=& -\frac{e^{2}}{V} \sum_{k,a,b}\sum_{k',a',b'}\int d\tau e^{i\Omega_{n}\tau}\partial_{k_{\alpha}} H_{\sigma}^{ab}(k)  \nonumber \\
	&\times& \partial_{k_{\alpha}'}H_{\sigma}^{a'b'}(k') G_{b' a}(k,-\tau)G_{b a'}(k,\tau)\delta_{k k'} \nonumber \\    
	&=& -\frac{e^{2}}{V} \sum_{k}\sum_{a,b,a',b'}\int d\tau e^{i\Omega_{n}\tau}\partial_{k_{\alpha}} H_{\sigma}^{ab}(k) \nonumber \\
	&\times&\partial_{k_{\alpha}}H^{a'b'}_{\sigma}(k) G_{b' a}(k,-\tau)G_{b a'}(k,\tau) \nonumber \\  
	&=& -\frac{e^{2}}{V} \sum_{k} \frac{1}{\beta}\sum_{\omega_{n}}\sum_{a,b,a',b'}\partial_{k_{\alpha}} H^{a' b'}_{\sigma}(k) \nonumber \\
	&\times& G_{b' a}(k,\omega_{n})\partial_{k_{\alpha}}H^{a b}_{\sigma}(k)G_{b a'}(k,\omega_{n}+\Omega_{n}) \nonumber \\ 
	&=&-\frac{e^2}{V}\sum_{k}\frac{1}{\beta}\sum_{\omega_{n}}\mathrm{Tr}[\partial_{k_{\alpha}}H_{\sigma}(k)G_{\sigma}(k,\omega_{n})\nonumber \\
	&\times& \partial_{k_{\alpha}}H_{\sigma}(k)G_{\sigma}(k,\omega_{n}+\Omega_{n})]. \nonumber
\end{eqnarray}

Then, exploiting the spectral function $G_{\sigma}(k,\omega_{n})=\int d\omega \frac{A_{\sigma}(k,\omega)}{i \omega_{n} - \omega}$, we can obtain
\begin{eqnarray}
	\Lambda_{\alpha \alpha}^{\sigma}(\Omega_{n})
	&=&-\frac{e^{2}}{V}\sum_{k}\frac{1}{\beta} \sum_{\omega_{n}} \mathrm{Tr}[\partial_{k_{\alpha}}H_{\sigma}(k)\int d \omega_{1} \frac{A_{\sigma}(k,\omega_{1})}{i \omega_{n}-\omega_{1}} \nonumber \\
	&\times&\partial_{k_{\alpha}}H_{\sigma}(k)\int d \omega_{2} \frac{A_{\sigma}(k,\omega_{2})}{i (\omega_{n}+\Omega_{n})-\omega_{2}}] \nonumber \\
	&=&-\frac{e^{2}}{V}\sum_{k} \int d \omega_{1} \int d \omega_{2} \frac{1}{\beta} \nonumber \\
	&\times&\sum_{\omega_{n}}\frac{1}{i \omega_{n}-\omega_{1}}\frac{1}{i (\omega_{n}+\Omega_{n})-\omega_{2}}   \nonumber \\
	&\times&\mathrm{Tr}[\partial_{k_{\alpha}}H_{\sigma}(k) A_{\sigma}(k,\omega_{1})\partial_{k_{\alpha}}H_{\sigma}(k) A_{\sigma}(k,\omega_{2})] \nonumber \\
	&=&-\frac{e^{2}}{V}\sum_{k} \int d \omega_{1} \int d \omega_{2} \frac{f_{F}(\omega_{1})-f_{F}(\omega_{2})}{i \Omega_{n}-\omega_{2}+\omega_{1}}   \nonumber \\
	&\times&\mathrm{Tr}[\partial_{k_{\alpha}}H_{\sigma}(k) A_{\sigma}(k,\omega_{1})\partial_{k_{\alpha}}H_{\sigma}(k) A_{\sigma}(k,\omega_{2})] 	  \nonumber 
\end{eqnarray}
where $f_{F}(x) = 1/(e^{x/T} + 1)$ is the standard Fermi distribution function. 

The above equation leads to the retarded current-current correlation by performing analytic continuation $i\Omega_{n}\rightarrow \omega + i 0^{+}$
\begin{eqnarray}
	\Lambda_{\alpha \alpha}(\omega )&=&-\frac{e^{2}}{V}\sum_{k} \int d \omega_{1} \int d \omega_{2} \frac{f_{F}(\omega_{1})-f_{F}(\omega_{2})}{\omega + i 0^{+}-\omega_{2}+\omega_{1}}   \nonumber \\
	&\times&\mathrm{Tr}[\partial_{k_{\alpha}}H_{\sigma}(k) A_{\sigma}(k,\omega_{1})\partial_{k_{\alpha}}H_{\sigma}(k) A_{\sigma}(k,\omega_{2})]. \nonumber 
\end{eqnarray}
Using the formula $\frac{1}{x+i 0^{+}}=\frac{1}{x}-i \pi \delta(x)$, we can easily obtain the imaginary part of the current-current correlation function
\begin{eqnarray}
	\mathrm{Im} \Lambda_{\alpha \alpha}^{\sigma}(\omega)&=&\frac{e^{2} \pi}{V}\sum_{k} \int d \omega_{1}  (f_{F}(\omega_{1})-f_{F}(\omega+\omega_{1}))\mathrm{Tr}[  \nonumber \\
	&~& \partial_{k_{\alpha}} H_{\sigma}(k) A_{\sigma}(k,\omega_{1})\partial_{k_{\alpha}}H_{\sigma}(k) A_{\sigma}(k,\omega+\omega_{1})]. \nonumber 
\end{eqnarray}
Therefore, the conductivity has the following formula  with considering the constant $\hbar$

\begin{eqnarray}
	\sigma_{\alpha \alpha}^{\sigma}&=&\frac{e^{2}}{\hbar}\frac{\pi}{V}\sum_{k} \int d \omega_{1}  (-\frac{d f_{F}(\omega_{1})}{d \omega_{1}})  \nonumber \\
	&\times&\mathrm{Tr}[\partial_{k_{\alpha}}H_{\sigma}(k) A_{\sigma}(k,\omega_{1})\partial_{k_{\alpha}}H_{\sigma}(k) A_{\sigma}(k,\omega_{1})]. \nonumber 
\end{eqnarray}  

The above formula can be used to calculate the conductivity for finite temperature. At zero temperature, we further simplify the evaluation by using $\frac{d f_{F}}{d \omega_{1}}=-\delta(\omega_{1}-\omega_{E_{F}})$.


\begin{thebibliography}{58}%

\bibitem{Smejkal2022} L. \v{S}mejkal, J. Sinova, T. Jungwirth, Emerging Research Landscape of Altermagnetism, Phys. Rev. X \textbf{12}, 040501 (2022).
\bibitem{PRXSmejkal2022} L. \v{S}mejkal, J. Sinova, T. Jungwirth, Beyond Conventional Ferromagnetism and Antiferromagnetism: A Phase with Nonrelativistic Spin and Crystal Rotation Symmetry, Phys. Rev. X \textbf{12}, 031042 (2022).
\bibitem{Mazin2022} I. I. Mazin, Editorial: Altermagnetism—A New Punch Line of Fundamental Magnetism, Phys. Rev. X \textbf{12}, 040002 (2022).

\bibitem{Ahn2019}
K.-H. Ahn, A. Hariki, K.-W. Lee and J. Kune\v{s},  Antiferromagnetism in RuO$_2$ as $d$-wave Pomeranchuk instability, Phys. Rev B \textbf{99}, 184432 (2019).
\bibitem{Phys.Soc.JpnHayami2019} S. Hayami, Y. Yanagi, and H. Kusunose, Momentum-Dependent Spin Splitting by Collinear Antiferromagnetic Ordering, J. Phys. Soc. Jpn. \textbf{88}, 123702 (2019).
\bibitem{Smejkal2020} L. \v{S}mejkal, R. Gonz\'{a}lez-Hern\'{a}ndez, T. Jungwirth, and J. Sinova, Crystal Time-Reversal Symmetry Breaking and Spontaneous Hall Effect in Collinear Antiferromagnets, Sci. Adv. \textbf{6}, eaaz8809 (2020).
\bibitem{PRBHayami2020} S. Hayami, Y. Yanagi, and H. Kusunose, Bottom-up design of spin-split and reshaped electronic band structures in antiferromagnets without spin-orbit coupling: Procedure on the basis of augmented multipoles, Phys. Rev. B \textbf{102}, 144441 (2020).
\bibitem{Yuan2020}
L.-D. Yuan, Z. Wang, J.-W. Luo, E. I. Rashba and A. Zunger, Giant momentum-dependent spin splitting in centrosymmetric low-$Z$ antiferromagnets, Phys. Rev. B \textbf{102}, 014422 (2020).
\bibitem{PRLGonzalez-Hernandez2021} R. Gonz\'{a}lez-Hern\'{a}ndez, L. \v{S}mejkal, K. V\'{y}born\'{y}, Y. Yahagi, J. Sinova, T. Jungwirth, and J. \v{Z}elezn\'{y}, Efficient Electrical Spin Splitter Based on Nonrelativistic Collinear Antiferromagnetism, Phys. Rev. Lett. \textbf{126}, 127701 (2021).
\bibitem{Mazin2021}
I. I. Mazin, K. Koepernik, M. D. Johannes, R. Gonz\'{a}lez-Hern\'{a}ndez and L. \v{S}mejkal, Prediction of unconventional magnetism in doped FeSb$_2$, Proc. Natl. Acad. Sci. U.S.A. \textbf{118}, e2108924118 (2021).
\bibitem{Bai2022}
H. Bai et al., Observation of Spin Splitting Torque in a Collinear Antiferromagnet RuO$_2$, Phys. Rev. Lett. \textbf{128}, 197202 (2022).


\bibitem{Bai2023}
H. Bai et al., Efficient Spin-to-Charge Conversion via Altermagnetic Spin Splitting Effect in Antiferromagnet RuO$_2$, Phys. Rev. Lett. \textbf{130}, 216701 (2023). 
\bibitem{PRBCui2023} Q. Cui, B. Zeng, P. Cui, T. Yu, and H. Yang, Efficient spin Seebeck and spin Nernst effects of magnons in altermagnets, Phys. Rev. B \textbf{108}, L180401 (2023).
\bibitem{AdvSciGuo2024}Y. Guo, et al., Direct and Inverse Spin Splitting Effects in Altermagnetic RuO$_2$, Adv. Sci., \textbf{11}, 2400967 (2024).
\bibitem{Sicheler2025}N. Sicheler, R. Raimondi, G. Sangiovanni, and  L. Del Re, Optically Tunable Spin Transport in Bilayer Altermagnetic Mott Insulators, arXiv:2508.06938.

\bibitem{PRLKarube2022} S. Karube, T. Tanaka, D. Sugawara, N. Kadoguchi, M. Kohda, and J. Nitta, Observation of Spin-Splitter Torque in Collinear Antiferromagnetic RuO$_2$, Phys. rev. lett. \textbf{129}, 137201 (2022).

\bibitem{Gonzalez2023}
R. D. Gonzalez Betancourt et al., Spontaneous Anomalous Hall Effect Arising from an Unconventional Compensated Magnetic Phase in a Semiconductor, Phys. Rev. Lett. \textbf{130}, 036702 (2023).
\bibitem{PRLSato2024} T. Sato, S. Haddad, I. C. Fulga, F. F. Assaad, and J. van den Brink, Altermagnetic anomalous Hall effect emerging from electronic correlations, Phys. Rev. Lett. \textbf{133}, 086503 (2024).
\bibitem{PRBAttias2024} L. Attias, A. Levchenko, and M. Khodas, Intrinsic anomalous Hall effect in altermagnets, Phys. Rev. B \textbf{110}, 094425 (2024).
\bibitem{NatCommunReichlova2024} H. Reichlova et al., Observation of a spontaneous anomalous Hall response in the Mn$_5$Si$_3$ $d$-wave altermagnet candidate, Nat. Commun. \textbf{15}, 4961 (2024).






\bibitem{Krempasky2024}
Krempask\'{y} et al., Altermagnetic lifting of Kramers spin degeneracy, Nature \textbf{626}, 517 (2024).
\bibitem{Fedchenko2024}
Fedchenko et al., Observation of time-reversal symmetry breaking in the band structure of altermagnetic RuO$_2$, Sci. Adv. \textbf{10}, eadj4883 (2024).

\bibitem{Feng2022}
Z. Feng et al., An anomalous Hall effect in altermagnetic ruthenium dioxide, Nat. Electron. \textbf{5}, 735 (2022).
\bibitem{Bhowal2024PRX} S. Bhowal and N. A. Spaldin, Phys. Ferroically Ordered Magnetic Octupoles in $𝑑$-Wave Altermagnets, Rev. X \textbf{14}, 011019 (2024).
\bibitem{Li2024PRB} Y.-X. Li, Y. Liu, and C.-C. Liu, Creation and manipulation of higher-order topological states by altermagnets, Phys. Rev. B \textbf{109}, L201109 (2024).
\bibitem{Hariki2024}
A. Hariki et al., X-Ray Magnetic Circular Dichroism in Altermagnetic $\alpha$-MnTe, Phys. Rev. Lett. \textbf{132}, 176701 (2024).
\bibitem{Lee2024}
S. Lee et al., Broken Kramers Degeneracy in Altermagnetic MnTe, Phys. Rev. Lett. \textbf{132}, 036702 (2024).
\bibitem{Osumi2024}
T. Osumi et al., Observation of a giant band splitting in altermagnetic MnTe, Phys. Rev. B \textbf{109}, 115102 (2024).

\bibitem{Yang2024}
G. Yang, Z. li, S. Yang et al., Three-dimensional mapping of the altermagnetic spin splitting in CrSb, Nat. Commun. \textbf{16}, 1442 (2025).



\bibitem{Brekke2023}
B. Brekke, A. Brataas and A. Sudb{\o}, Two-dimensional altermagnets: Superconductivity in a minimal microscopic model, Phys. Rev. B \textbf{108}, 224421 (2023).
\bibitem{Mland2024}
K. M{\ae}land, B. Brekke and A. Sudb{\o},  Many-body effects on superconductivity mediated by double-magnon processes in altermagnets, Phys. Rev. B \textbf{109}, 134515 (2024).
\bibitem{Bose2024}
A. Bose, S. Vadnais and A. Paramekanti, Altermagnetism and superconductivity in a multiorbital t-J model, Phys. Rev. B \textbf{110}, 205120 (2024).
\bibitem{Maier2023}
T. A. Maier and S. Okamoto, Weak-coupling theory of neutron scattering as a probe of altermagnetism, Phys. Rev. B \textbf{108}, L100402 (2023).
\bibitem{Das2024} P. Das, V. Leeb, J. Knolle and M. Knap, Realizing Altermagnetism in Fermi-Hubbard Models with Ultracold Atoms,  Phys. Rev. Lett. \textbf{132}, 263402 (2024).
\bibitem{Leeb2024}
V. Leeb, A. Mook, L. \v{S}mejkal and J. Knolle, Spontaneous Formation of Altermagnetism from Orbital Ordering, Phys. Rev. Lett. \textbf{132}, 236701 (2024).

\bibitem{Antonenko2025} D. S. Antonenko, R. M. Fernandes, and J. W. F. Venderbos, Mirror Chern Bands and Weyl Nodal Loops in Altermagnets, Phys. Rev. Lett. \textbf{134}, 096703 (2025).
\bibitem{DelRe2025} L. Del Re, Dirac points and topological phases in correlated altermagnets, Phys. Rev. Res. \textbf{7}, 033234 (2025). 

\bibitem{Zhu2023}
D. Zhu, Z.-Y. Zhuang, Z. Wu and Z. Yan, Topological superconductivity in two-dimensional altermagnetic metals, Phys. Rev. B \textbf{108}, 184505 (2023).
\bibitem{PRLOuassou2023} J. A. Ouassou, A. Brataas, and J. Linder, Dc josephson effect in altermagnets, Phys. Rev. Lett. \textbf{131}, 076003 (2023).
\bibitem{Chakraborty2023}
D. Chakraborty and A. M. Black-Schaffer, Zero-field finite-momentum and field-induced superconductivity in altermagnets, Phys. Rev. B \textbf{110}, L060508 (2024).
\bibitem{PRBFernandes2024} R. M. Fernandes, V. S. de Carvalho, T. Birol, and R. G. Pereira, Topological transition from nodal to nodeless Zeeman splitting in altermagnets, Phys. Rev. B \textbf{109}, 024404 (2024).
\bibitem{PRBCheng2024} Q. Cheng and Q.-F. Sun, Orientation-dependent Josephson effect in spin-singlet superconductor/altermagnet/spin-triplet superconductor junctions, Phys. Rev. B \textbf{109}, 024517 (2024).
\bibitem{PRBDiniz2024} G. S. Diniz and  E. Vernek, Suppressed Kondo screening in two-dimensional altermagnets, Phys. Rev. B \textbf{109}, 155127 (2024).
\bibitem{ArxivLee2023}Y.-L. Lee, Magnetic impurities in an altermagnetic metal, Eur. Phys. J. B \textbf{98}, 43 (2025).


\bibitem{Hewson1993} A.Hewson, The Kondo Problem to Heavy Fermions(Cambridge University Press, Cambridge, 1993).
\bibitem{arXivHe2024} X. He and S. Zhang, Dirac fermions in the altermagnet Ce$_4$Sb$_3$, Phys. Rev. B \textbf{112}, 075138 (2025).
\bibitem{arXivHellenes2023} A. B. Hellenes, T. Jungwirth, R. Jaeschke-Ubiergo, A. Chakraborty, J. Sinova, and L. Smejkal, $P$-wave magnets, arXiv:2309.01607.
\bibitem{PRBAmundsen2024} M. Amundsen, A. Brataas, and J. Linder, RKKY interaction in Rashba altermagnets, Phys. Rev. B \textbf{110}, 054427 (2024).


\bibitem{Tsunetsugu} H. Tsunetsugu, M. Sigrist and K. Ueda, The ground-state phase diagram of the one-dimensional Kondo lattice model, Rev. Mod. Phys. \textbf{69}, 809 (1997).
\bibitem{Coleman2015} P. Coleman, Introduction to Many Body Physics (Cambridge University Press, 2015).


\bibitem{Lacroix1979} C. Lacroix and M. Cyrot, Phase diagram of the Kondo lattice, Rev. Mod. B \textbf{20}, 1969 (1979).

\bibitem{Fazekas1991}
P. Fazekas and E. M\"{u}ller-Hartmann, Magnetic and non-magnetic ground states of the Kondo lattice, Z. Phys. B  \textbf{85}, 285 (1991).

\bibitem{Zhang2000}
G.-M. Zhang and L. Yu, Kondo singlet state coexisting with antiferromagnetic long-range order: A possible ground state for Kondo insulators, Phys. Rev. B \textbf{62}, 76 (2000).

\bibitem{Li2015}
H. Li, Y. Liu, G.-M. Zhang and L. Yu, Phase evolution of the two-dimensional Kondo lattice model near half-filling, J. Phys.: Condens. Matter \textbf{27}, 425601 (2015).


\bibitem{Watanabe2007}
H. Watanabe and M. Ogata, Fermi-Surface Reconstruction without Breakdown of Kondo Screening at the Quantum Critical Point, Phys. Rev. Lett. \textbf{99}, 136401 (2007).
\bibitem{Asadzadeh2013}
M. Z. Asadzadeh, F. Becca and M. Fabrizio, Variational Monte Carlo approach to the two-dimensional Kondo lattice model, Phys. Rev. B \textbf{87}, 205144 (2013).
\bibitem{Martin2008}
L. C. Martin and F. F. Assaad, Evolution of the Fermi Surface across a Magnetic Order-Disorder Transition in the Two-Dimensional Kondo Lattice Model: A Dynamical Cluster Approach, Phys. Rev. Lett. \textbf{101}, 066404 (2008).
\bibitem{Lenz2008}
B. Lenz, R. Gezzi and S. R. Manmana, Variational cluster approach to superconductivity and magnetism in the Kondo lattice model, Phys. Rev. B \textbf{96}, 155119 (2017).

\bibitem{Zhong2024} M. Zhao, W.-W. Yang, X. Guo, H.-G. Luo, Y. Zhong, Altermagnetism in Heavy Fermion Systems, Phys. Rev. B \textbf{111}, 085145 (2025).

\bibitem{Assaad1999}
F. F. Assaad, Quantum Monte Carlo Simulations of the Half-Filled Two-Dimensional Kondo Lattice Model, Phys. Rev. Lett. \textbf{83}, 796 (1999).


\bibitem{Lohneysen2007}H. v. Löhneysen, A. Rosch, M. Vojta, and P. Wölfle, Fermi-liquid instabilities at magnetic quantum phase transitions, Rev. Mod.
Phys. \textbf{79}, 1015 (2007).
\bibitem{Si2013} Q. Si and S. Paschen, Quantum phase transitions in heavy fermion metals and Kondo insulators, Phys. Status Solidi B \textbf{250}, 425 (2013).
\bibitem{Coleman2010}P. Coleman and A. H. Nevidomskyy, Frustration and the Kondo Effect in Heavy Fermion Materials, J. Low Temp. Phys. \textbf{161}, 182 (2010).

\bibitem{PRBSikkema1996} A. E. Sikkema, W. J. L. Buyers, I. Affleck, and J. Gan, Ising-Kondo lattice with transverse field: A possible f-moment Hamiltonian for URu$_2$Si$_2$, Phys. Rev. B \textbf{54}, 9322 (1996).
\bibitem{RevModPhysMydosh2011} J. A. Mydosh and P. M. Oppeneer, Colloquium: Hidden order, superconductivity, and magnetism: The unsolved case of URu$_2$⁢Si$_2$, Rev. Mod. Phys. \textbf{83}, 1301 (2011).

\bibitem{PRBYang2019} W.-W. Yang, J. Zhao, H.-G. Luo, and Y. Zhong, Exactly solvable Kondo lattice model in the anisotropic limit, Phys. Rev. B \textbf{100},  045148 (2019).
\bibitem{PRBYang2020}W.-W. Yang, Y. Zhong, and H.-G. Luo, Hexagonal Ising-Kondo lattice: An implication for intrinsic antiferromagnetic topological insulator, Phys. Rev. B \textbf{102}, 195141(2020).
\bibitem{PRBYang2021} W.-W. Yang, Y.-X. Li,  Y. Zhong, and H.-G. Luo, Doping a Mott insulator in an Ising-Kondo lattice: Strange metal and Mott criticality, Phys. Rev. B \textbf{104}, 165146 (2021).


\bibitem{Maska2006} M. M. Maska and K. Czajka, Thermodynamics of the two-dimensional Falicov-Kimball model: A classical Monte Carlo study, Phys. Rev. B \textbf{74}, 035109 (2006).
\bibitem{PRLFalicov1969} L. M. Falicov and J. C. Kimball, Simple Model for Semiconductor-Metal Transitions: SmB$_6$ and Transition-Metal Oxides, Phys. Rev. Lett. \textbf{22}, 997 (1969).
\bibitem{PhysicaAKennedy1986} T. Kennedy and E. H. Lieb, An itinerant electron model with crystalline or magnetic long range order, Physica A \textbf{138}, 320 (1986).

\bibitem{PRBChen2024} W. Chen, X. Zhou, D. Zhang, Y.-Q. Xu, and W.-K. Lou, Impurity scattering and Friedel oscillations in altermagnets, Phys. Rev. B \textbf{110}, 165413 (2024).
\bibitem{PRBSukhachov2024} P. Sukhachov and J. Linder, Impurity-induced Friedel oscillations in altermagnets and $p$-wave magnets, Phys. Rev. B \textbf{110}, 205114 (2024).





































\end{thebibliography}
\end{document}